\documentclass[12pt]{iopart}

\usepackage{graphicx, amssymb, amsfonts, cite, braket, color}
\newcommand{\hH}{\hat{H}}
\newcommand{\hs}{\hat{\sigma}}
\newcommand{\hz}{\hat{z}}
\newcommand{\hx}{\hat{x}}
\newcommand{\I}{\mathcal{I}}
\newcommand{\tk}{\tilde{k}}
\newcommand{\tC}{\tilde{C}}
\newcommand{\td}{\tilde{d}}

\begin{document}

\title{Response of quantum spin networks to attacks}

\author{Bhuvanesh Sundar$^{1,2,3}$ Mattia Walschaers$^4$, Valentina Parigi$^4$, and Lincoln D Carr$^5$}
\address{$^1$ Institute for Quantum Optics and Quantum Information of the Austrian Academy of Sciences, Innsbruck A 6020, Austria}
\address{$^2$Department of Physics and Astronomy, Rice University, Houston, Texas 77005, USA}
\address{$^3$Rice Center for Quantum Materials, Rice University, Houston, Texas 77005, USA}
\address{$^4$Laboratoire Kastler Brossel, Sorbonne Universit{\'e}, CNRS, ENS-PSL Research University, Coll{\'e}ge de France, Paris F-7252, France}
\address{$^5$Department of Physics, Colorado School of Mines, Golden, Colorado 80401, USA}
\eads{\mailto{bhuvanesh.sundar@colorado.edu}, \mailto{mattia.walschaers@lkb.upmc.fr}, \mailto{valentina.parigi@lkb.upmc.fr}, \mailto{lcarr@mines.edu}}

\begin{abstract}
We investigate the ground states of spin models defined on networks that we imprint (e.g., non-complex random networks like Erdos-Renyi, or complex networks like Watts-Strogatz, and Barabasi-Albert), and their response to decohering processes which we model with network attacks. We quantify the complexity of these ground states, and their response to the attacks, by calculating distributions of network measures of an emergent network whose link weights are the pairwise mutual information between spins. We focus on attacks which projectively measure spins. We find that the emergent networks in the ground state do not satisfy the usual criteria for complexity, and their average properties are captured well by a single dimensionless parameter in the Hamiltonian. While the response of classical networks to attacks is well-studied, where classical complex networks are known to be more robust to random attacks than random networks, we find counter-intuitive results for our quantum networks. We find that the ground states for Hamiltonians defined on different classes of imprinted networks respond similarly to all our attacks, and the attacks rescale the average properties of the emergent network by a constant factor. Mean field theory explains these results for relatively dense networks, but we also find the simple rescaling behavior away from the regime of validity of mean field theory. Our calculations indicate that complex spin networks are not more robust to projective measurement attacks, and presumably also other quantum attacks, than non-complex spin networks, in contrast to the classical case. Understanding the response of the spin networks to decoherence and attacks will have applications in understanding the physics of open quantum systems, and in designing robust complex quantum systems -- possibly even a robust quantum Internet in the long run -- that is maximally resistant to decoherence.
\end{abstract}

\section{Introduction}

The textbook example of a quantum wave function is that of an isolated quantum system in the ground state of a Hamiltonian. Experimental advances in quantum technology have led researchers to reliably realize this textbook example in labs. Nowadays, experimentalists routinely create strongly correlated quantum states with tens to hundreds of spins or qubits in the ground state of interacting quantum Hamiltonians, for applications in quantum simulation, computation and communication~\cite{altman2021quantum,awschalom2021development,alexeev2021quantum}.

A significant challenge in these experiments, after creating the desired quantum state, is preserving the state against decoherence. Decoherence occurs because the quantum system is not completely isolated, but interacts with the environment in several ways. Examples of decoherence include motional heating, atom losses, spin dephasing, depolarizing, and spontaneous emissions. We investigate the effects of dephasing on ground states of quantum spin models defined on imprinted spin networks.

We use a network-science approach to represent the quantum spin system and to capture dephasing effects on network properties. Classical network science is a well-developed field which studies complex networks such as telecommunication networks, computer networks, biological networks, social networks, and cognitive and semantic networks~\cite{newman2018networks}. Complex networks have been shown to be more vulnerable to some network attacks than others, and more robust to certain attacks than random and other non-complex networks. These phenomena have been studied in the classical context, i.e., classical attacks occurring on classical networks. Often, such attacks involve nodes being removed from the network.

Quantum complex networks is a much younger field with many basic open questions to be resolved~\cite{Bianconi2001,acin2007entanglement,plenio2008dephasing,faccin2013degree,bianconi2015complex,garcia2020pairwise}. Quantum complex networks can be either imprinted in the Hamiltonian, quantum circuit, or quantum architecture~\cite{Bianconi2001,acin2007entanglement,plenio2008dephasing,faccin2013degree}; or they can be emergent in the quantum state~\cite{valdez2017quantifying,sundar2018complex,bagrov2019detecting,garcia2020pairwise}. Attacks may take a variety of forms, from node removal to weak or strong measurement, either randomly or preferentially. Among these many unanswered questions in quantum complex networks, we consider the most pressing one, namely the effects of decoherence modeled via projective measurement, construed as a quantum analogy of classical network attacks.

We consider the system to have spins arranged on the nodes of a complex network (also called a graph), with the interactions between the spins determined by their arrangement on the network. We calculate emergent properties in the ground state of a Hamiltonian defined on this network. We argue that the emergent properties give rise to an \emph{emergent network}. Specifically, we construct the emergent network with spins as nodes and link weights given by the pairwise mutual information (MI). This approach has had success in past contexts from quantum phase transitions~\cite{valdez2017quantifying} to quantum cellular automata~\cite{hillberry2020entangled}. We calculate the MI across a range of Hamiltonian parameters. Then, we attack the networks as we describe below.

A common model of decoherence in quantum computing is probabilistic projective attacks in a quantum trajectories approach, also used for quantum simulations with many-body systems~\cite{daley2014quantum} such as the spin network considered here. This kind of network attack can be expected to cause pair-wise entanglement to rapidly decay, as in Google's quantum supremacy experiment on random states~\cite{arute2019quantum}.
Thus, we focus on the case that attacks projectively measure the spins, and vary the strength and direction of these measurements, as well as different strategies for choosing the attacked spins -- randomly or preferentially -- analogous to classical attacks on classical networks. We quantify the effect of these attacks on the ground state by calculating network measures of the emergent MI network before and after the attacks. These attacks are not normally treated in decoherence models, but are of interest to the fundamental theory of quantum complex networks as compared to classical ones.

Previous works have studied quantum systems with a complex network structure imprinted on them~\cite{Bianconi2001, Bianconi2012a, Bianconi2012b, Kimble2008, Perseguers2010, Cuquet2009, acin2007entanglement}, while others have shown that the emergent MI network in the state is complex~\cite{valdez2017quantifying, sundar2018complex, zaman2019real}. But these works did not investigate the consequences of complexity in these quantum systems. By considering the effects of attacks on quantum networks, we address the important question of the vulnerability of imprinted complex networks to attacks, and more broadly, to decoherence. Understanding this vulnerability has implications in designing robust quantum systems, and ultimately, both a robust quantum internet and a robust networking choice for separate NISQ-era computers~\cite{awschalom2021development}.

\subsection{Overview of the Article}

In Sec.~\ref{sec: Imprinted vs. Emergent Quantum Complex Networks}, we overview the wide variety of quantum complex networks in contrast to classical ones; describe our choice of imprinted networks in Hamiltonians; and explain the emergent MI network in the resulting quantum many-body ground states. In Sec.~\ref{sec: Hamiltonian}, we describe the specific Hamiltonian we consider, and the numerical and analytical tools we use to study the system. In Sec.~\ref{sec: MI ground state}, we apply complex network analysis to the emergent MI network in the ground state. In Sec.~\ref{sec: attacks}, we overview attacks on quantum complex networks in general and describe in detail the attacks we consider here. In Sec.~\ref{sec: results}, we determine the response of the MI networks to attack, and show how mean-field (MF) theory captures many of the results, independent of imprinted network choice. We summarize in Sec.~\ref{sec: summary}.

Our main results are summarised below:
\begin{enumerate}
\item The emergent MI networks in the ground state do not satisfy the usual criteria for complexity, such as a power-law degree distribution or high clustering. Instead, for relatively dense imprinted networks, the average properties of the emergent networks are captured well by mean-field (MF) theory which predicts fully-connected emergent networks. MF theory does not capture the higher moments of these properties, and also fails to capture the emergent network properties for sparse imprinted networks.
\item The lowest order moment of properties of the emergent MI network, for different dense imprinted networks, overlap with each other when plotted versus a dimensionless parameter in the problem.
\item The above results hold both for the ground state and the attacked system.
\item The average properties of the emergent MI network before and after the attack also collapse onto each other after an appropriate rescaling. Thus, the emergent MI network in the ground state of an imprinted complex network shows the same amount of robustness to projective measurement attacks as does that in the ground state of an imprinted random non-complex network.
\item For sparse imprinted networks, the lowest order moment of emergent MI network's properties do not overlap with each other for different classes of networks. However, the average properties of the emergent MI network before and after the attack again collapse onto each other after an appropriate rescaling.
\end{enumerate}
All these behaviors are unexpected, and completely different from the case of classical attacks on networks, which are known to depend strongly on the kind of network: random networks quickly fall apart, while complex networks maintain their properties and are therefore robust. We show that emergent quantum complex networks do \emph{not} share these robustness properties: (1) the choice of network structure does not help with decoherence; and (2) targeted decoherence is no more effective than random decoherence. We explain these phenomena with MF theory.

\section{Background}\label{sec: Imprinted vs. Emergent Quantum Complex Networks}

In this section, we first give an overview of quantum complex networks in Sec.~\ref{subsec: qtm vs classical}, and highlight how they can differ from classical complex networks. Then, in Sec.~\ref{subsec: imprinted network}, we describe the networks we \textit{imprint} on the Hamiltonian. In Sec.~\ref{subsec: MI}, we describe the network we study, which is the network that \textit{emerges} in the ground state. In Sec.~\ref{subsec: measures}, we describe the complex network measures, which are the tools we use to study the emergent network.

\subsection{Quantum vs classical complex networks} \label{subsec: qtm vs classical}

A network is an abstract representation of connections (links) between agents (nodes). Networks are nearly everywhere around us, and have a variety of structures. Complex networks are a subset of networks with an emergent property, such as a non-trivial degree distribution or a short path length. Complex networks occur in many classical scenarios, ranging from telecommunication networks to computer networks, biological networks, social networks, and cognitive and semantic networks~\cite{newman2018networks}. There are many models to generate networks, both complex and non-complex, that reproduce different properties of real-world networks such as those mentioned above. We will consider three of these models in Sec.~\ref{subsec: imprinted network}.

We now turn our attention to quantum networks. Quantum networks have been considered in at least three different settings. The first and most popular setting for quantum networks is the case where the links are physical, i.e., spins or qubits interact across either naturally occurring or engineered links~\cite{cirac1997quantum, chaneliere2005storage, wilk2007single, politi2008silica, ritter2012elementary, aspuru2012photonic}. In this work, we refer to this kind of network as the \textit{imprinted} network. The second scenario where quantum networks have been studied is the case of networks whose links are represented by entangled states~\cite{acin2007entanglement, Cuquet2009, Perseguers2010}. Loosely speaking, this network is defined on the quantum many-body state. In this work, we refer to this kind of network as the \textit{emergent} network. A third setting is the case where the network is defined on only one part of the Hamiltonian, e.g. via perturbation theory. This third scenario has already been explored in experiments on electronic transport through quantum antidots~\cite{Poteshman_2019}; and in an open quantum system the system can be simple with a complex environment, or vice-versa~\cite{nokkala2016complex}.

In this work, we study the complex network measures of the emergent network that exists in the quantum many-body ground state of a Hamiltonian defined on an imprinted network. We will describe the imprinted and emergent networks in detail in the following sections. We emphasize that the emergent network can have a very different structure from the imprinted network: complexity in the Hamiltonian may or may not give rise to complexity in the state, just as complexity can also arise out of non-complex simple systems. This is a non-trivial question that has to be addressed on a case-by-case basis, as we have separately considered for continuous-variable multimode quantum networks~\cite{Walschaers2020}. The general question of the relationship between complexity and emergence remains an open one~\cite{thurner2018introduction}.

While previous studies have focused on using quantum networks to, for example, establish quantum communication over long distances, we focus on the complex networks properties of the emergent network in the quantum many-body ground state defined on imprinted networks. Our work thus opens a new avenue for using network science to study quantum systems.

\subsection{Imprinted Spin Networks} \label{subsec: imprinted network}

Hamiltonians are usually defined on regular lattices with regular links in condensed matter and statistical physics, and have random links in random matrix theory, e.g. in nuclear physics or spin glasses in condensed matter physics. Our novelty is that we generalize these Hamiltonians to complex networks. Specifically, we define the Hamiltonian as $\hH = \sum_{\braket{ ij}} \hH_{ij}$, where the sum runs over the links in the network. We call the network with these links $\hH_{ij}$ as the imprinted network. Hamiltonians defined on such non-regular imprinted networks have much more complex phases such as spin glasses already at the classical level, and also form the basis of Hopfield networks for artificial neural nets. The phase diagram of Hamiltonians defined on networks which are neither regular nor random is an unexplored frontier of research.

We consider imprinted networks generated from three different models -- the Erdos-Renyi (ER) model, the Watts-Strogatz (WS) model, and the Barabasi-Albert (BA) model. The structure of the networks generated from these models differ significantly from the case of regular lattices. The ER model is random. The WS model is a small-world network, while the BA model shows power-law scaling and is thus called scale-free. The WS and BA models are therefore referred to as complex, in contrast to the ER model or regular lattices. 

Networks in each of the above models are constructed and parameterised differently, as detailed below. In this paper, we will find that an important parameter describing the universal physics of the ground state of $\hH$ is the average degree of each node in the imprinted network, $Z = \sum_i \tk_i/n$, also called the coordination number, where $\tk_i$ is the degree of node $i$.

Networks generated by the ER model are parameterized by a number $0\leq p\leq 1$, which specifies the probability that any pair of nodes are connected by a link. The distribution of the degree $\tk_i$ of node $i$ is binomial, and the average degree is peaked around $Z = (n-1)p$.

Networks generated by the WS model are parameterized by two numbers, an even integer $K$ which specifies the mean degree, and $0\leq p\leq 1$. Here, $K$ is the mean degree of a node, and $p$ is the probability that a link between two nodes $(i,j)$ is removed and rewired to connect $i$ to another node $j'$. When $p=0$, WS networks are regular, and when $p=1$, they are random and have a structure close to ER networks~\footnote{An ensemble of WS networks at $p=1$ still has some differences with an ensemble of ER networks. For example, the minimum possible degree of any node in any WS network is $K/2$, and the total number of links is $n*K/2$. Whereas, the minimum possible degree of any node in a connected ER network is $1$, and the total number of links is binomially distributed.}.

Networks generated by the BA model are parameterized by an integer $m$. The BA model is a preferential attachment model in which every new node is attached to $m$ existing nodes of the network, with probability of the new node attaching to an existing node $i$ given by $p_i=\tk_i/\sum_j \tk_j$. The BA model is scale-free, because upon adding nodes, it quickly generates a power-law degree distribution of $P(k)\approx k^{-3}$. The average degree of a BA network is fixed, $Z = 2m(n-m)/n$.

\begin{figure}[t]\centering
\includegraphics[width=0.6\columnwidth]{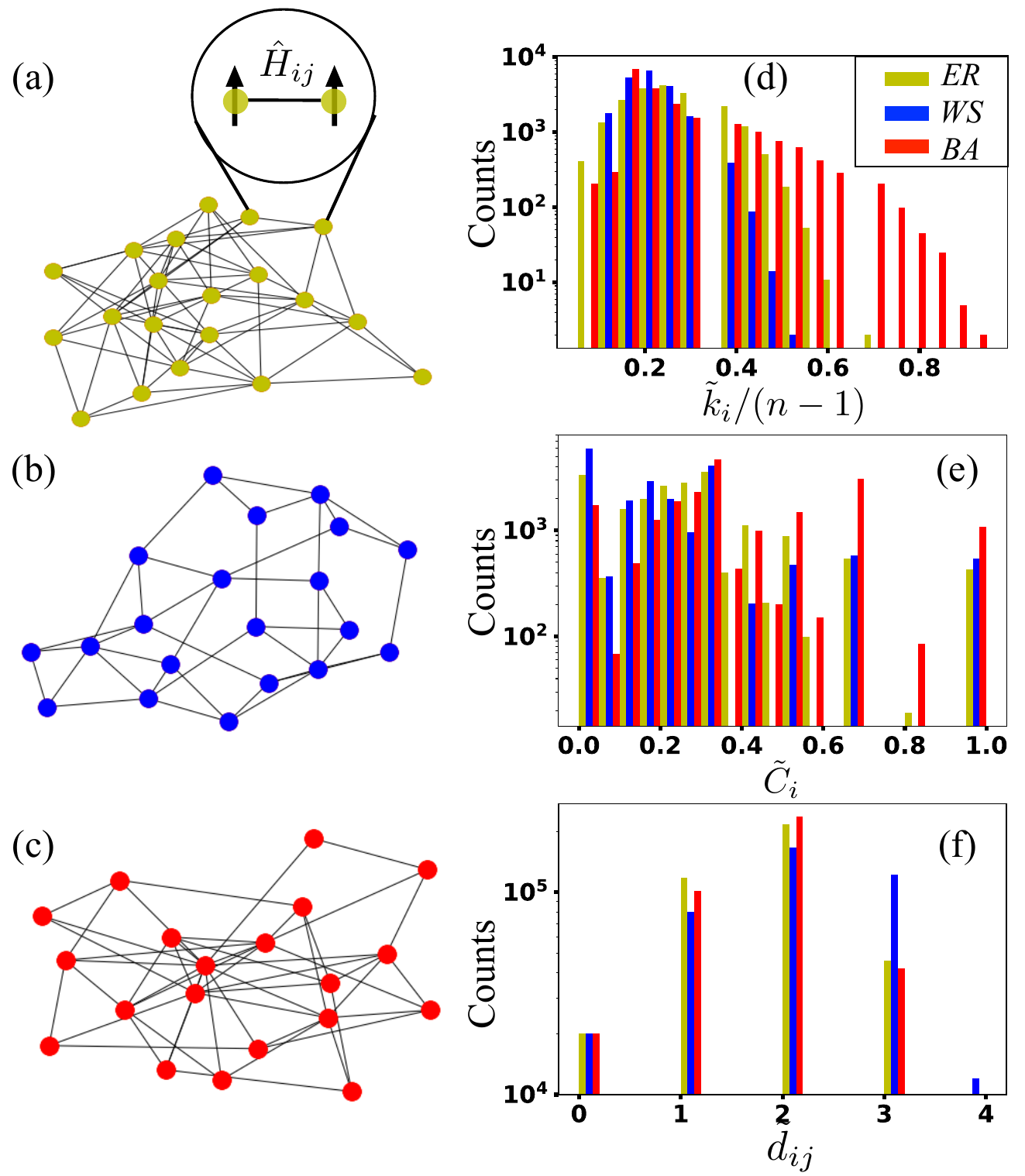}
\caption{Examples of networks with $n=20$ nodes drawn from (a) the ER model with $p=0.26$, (b) the WS model with $K=4$ and $p=0.5$, and (c) the BA model with $m=2$. (d-f) Distributions of the degree $\tk_i$, clustering $\tC_i$, and path lengths $\td_{ij}$ for an ensemble of $1000$ imprinted networks with parameters as in (a-c). Yellow bars show the distributions for ER networks, blue bars for WS, and red bars for BA networks.
}
\label{fig1}
\end{figure}

Figures~\ref{fig1}(a-c) show examples of networks generated from the ER, WS, and BA models.

\subsection{Emergent networks} \label{subsec: MI}

Central to our work is the emergent network that we define in the ground state of $\hH$. This emergent network is a fully connected network, whose nodes are the same as those of the imprinted network, and links are given by the quantum mutual information $\I_{ij}$ between spins $i$ and $j$. Our main focus in this work is analyzing the emergent network, before and after network attacks, and quantifying the complexity of $\I_{ij}$ via complex network measures defined in Sec.~\ref{subsec: measures}

The quantum mutual information $\I_{ij}$ between spins $i$ and $j$ is constructed from the one- and two-point von Neumann entropies $S_i = -\Tr(\rho_i\log\rho_i)$ and $S_{ij}= -\Tr(\rho_{ij}\log\rho_{ij})$ as
\begin{equation}\label{eqn: Iij}
\I_{ij} = \frac{1}{2}\left( S_i+S_j-S_{ij}\right),
\end{equation}
where $\rho_i=\Tr_{k\neq i}(\rho)$ and $\rho_j=\Tr_{k\neq j}(\rho)$ are the reduced density matrices for the spins labeled $i$ and $j$ in the quantum many-body ground state, and $\rho_{ij}=\Tr_{k\neq i,j}(\rho)$ is the reduced density matrix for the subsystem containing only the two spins $i$ and $j$. We define $\I_{ii}=0\ \forall i$.

While the quantum mutual information does not fully describe the system (a full description is given by the density matrix), there are still several motivations for using quantum mutual information. First, the pairwise MI is an upper bound to the squares of all two-body correlations along any direction~\cite{wolf2008area}. This can be important since one is often only interested in knowing two-body correlations.  Second, the MI network captures a lot of the richness of the system's quantum wave function, including long-range correlations and entanglement. Notably, previous works have shown that the variation of MI across the phase diagram mirrors the system's underlying phases~\cite{sundar2018complex, valdez2017quantifying, zaman2019real,bagrov2019detecting}. Third, the MI is a direct quantum analog of the most common complex network measures used in classical networks for EEG and fMRI~\cite{bullmore2009complex}, and is therefore an established choice from classical complex network theory and well understood. Finally, the MI taken as a complex network has proven useful in a number of quantum contexts, including complex dynamics~\cite{hillberry2020entangled}.

\subsection{Complex Network Measures} \label{subsec: measures}

Visualizing the full emergent MI network is cumbersome, and even more so when we consider an ensemble of imprinted networks. Therefore, we distill the MI in the emergent network by a few network measures. These are the degree $k_i$ and clustering coefficient $C_i$ on a node $i$, and shortest path length $d_{ij}$ between $i$ and $j$, generalized to analyze weighted networks as follows:
\begin{equation}\begin{array}{lcl}
k_i &=& \sum_j \I_{ij},\\
C_i &=& \left( \sum_{j\neq k} \I_{ij} \I_{jk} \I_{ki} \right) / \left( \sum_{j\neq k} \I_{ij} \I_{ik} \right),\\
d_{ij} &=& {\rm min}_{i_1,i_2,\cdots i_n} P(i\rightarrow i_1\rightarrow i_2 \rightarrow \cdots \rightarrow j),
\end{array}\end{equation}
where $P(i\rightarrow i_1\rightarrow i_2 \rightarrow \cdots \rightarrow j) = 1/\I_{ii_1} + 1/\I_{i_1i_2}+\cdots+\I_{i_nj}$ is the length of the path from $i$ to $j$ via the nodes $(i_1, i_2, \cdots i_n)$. Here, we have defined $1/\I_{ij}$ as the distance between two nodes $i$ and $j$ on the MI network via a direct link connecting them. Other network measures, such as disparity, geodesic distances, and various centrality measures can also yield useful information about the network~\cite{sundar2018complex}. Here we focus on $k_i$, $C_i$, and $d_{ij}$. We denote the corresponding measures for the imprinted networks as $k_i$, $C_i$, and $d_{ij}$, without the tilde.

The complex network measures also help us in classifying a network as complex or not. Complexity is not uniquely defined, and a working definition is often assumed to be a non-trivial topological feature such as a power-law degree distribution, or a higher clustering and shorter path length than random networks with the same average degree. In the following sections, we will use these working definitions to classify our emergent networks as complex or otherwise.

To illustrate the usefulness of these complex network measures, we briefly return to the imprinted networks. Figures~\ref{fig1}(d-f) show the distributions of $\tk_i$, $\tC_i$, and $\td_{ij}$ for an ensemble of imprinted ER, WS and BA networks. We clearly see that BA networks have a longer tail for $\tk_i$ than ER and WS networks, showing a hint of the power law decay already for $n=20$. The BA networks also have a shorter path length on average than WS and ER networks.

\section{Setup}\label{sec: Hamiltonian}

The specific Hamiltonian we consider on the imprinted network is one where spins across links have a $z$-$z$ interaction, and each spin experiences a transverse magnetic field along the $x$ direction. The Hamiltonian for the system is
\begin{equation} \label{eqn: H}
\hH = -\sum_{\braket{ ij}} J\hs^z_i \hs^z_j + h \sum_i \hs^x_i.
\end{equation}

This Hamiltonian is well-studied in the context of regular lattices. In this case, the Hamiltonian is called the transverse Ising model (TIM). It is one of the simplest exactly solvable quantum models in one dimension (1D), and has a well-understood phase diagram. In 1D chains, the TIM has a ferromagnetic ground state for $|h|<J$ and a paramagnetic ground state for $|h|>J>0$. Despite its simplicity, recent studies have shown that thermal states in the 1D TIM show complexity in emergent weighted networks whose edge weights are the von Neumann MI between particles~\cite{sundar2018complex}. The phase diagram of Eq.~(\ref{eqn: H}) on ER, WS and BA networks is not known.

\subsection{Analytical and numerical tools}\label{subsec: MF}
We exactly diagonalize $\hH$ [Eq.~(\ref{eqn: H})] to find its ground state, and calculate $\I_{ij}$ and the measures defined on the MI network. For each imprinted network model, we consider an ensemble of $100$ imprinted networks, and find moments of the network measures in this ensemble. We are limited to imprinted networks with $20$ nodes, since the computations become exponentially harder for larger networks.

To support and understand our exact diagonalization results, we also calculate the ground state of Eq.~(\ref{eqn: H}) within MF theory. This theory, widely applied in condensed matter and statistical physics, works by assuming that each particle feels its immediate neighborhood in the system as a mean field. In practice, this is done by approximating terms in the Hamiltonian by complex scalars, and self-consistently or variationally finding these scalars. It typically works well in systems with a large dimension $D$, i.e., particles arranged in a regular lattice with $Z=2D$. We find that it is sufficient to have large $Z$ for MF theory to be valid, and the regular lattice is not necessary. We show that MF theory accurately predicts the average properties of the MI network for dense imprinted random and complex networks, i.e. those with $Z\gtrsim4$.

To derive the MI, we make the MF approximation that each spin lies in the $x$--$z$ plane, and points at an angle $\theta$ to the $z$-axis. For simplicity, we first consider the case that all spins have the same $Z$. Then, the expectation value of Eq.~(\ref{eqn: H}) is
\begin{equation} \label{eqn: HMF}
\braket{ \hH } = -JZN \left( \frac{1}{2}\cos^2\theta + \lambda\sin\theta \right),
\end{equation}
where $\lambda = h/(ZJ)$. We solve for $\theta$ by minimizing Eq.~(\ref{eqn: HMF}). We obtain two solutions for $\lambda<1$, $\theta = \sin^{-1}\lambda$ and $\theta=\pi-\sin^{-1}\lambda$, and one solution for $\lambda>1$, $\theta=\pi/2$. Then the MF ground state is
\begin{eqnarray}\label{eqn: MF}
\ket{\psi} = \frac{ \ket{mm\cdots} + \ket{-\!m -\!m\cdots} }{\sqrt{2(1+\sin^n\theta)}} \quad {\rm for}\ \lambda<1, \nonumber\\
\ket{\psi} = \ket{\rightarrow\rightarrow\cdots} \quad {\rm for}\ \lambda>1,
\end{eqnarray}
where $m = \sqrt{1-\lambda^2}\Theta(1-\lambda)$, $\ket{m} = \cos\frac{\theta}{2}\ket{\uparrow} + \sin\frac{\theta}{2}\ket{\downarrow}$, and $\ket{-\!m} = \sin\frac{\theta}{2}\ket{\uparrow} + \cos\frac{\theta}{2}\ket{\downarrow}$. Note that $\ket{m}$ and $\ket{-\!m}$ are not orthogonal, and $\braket{ m\vert -\!m} = \sin\theta$, which explains the denominator on the first line of Eq.~(\ref{eqn: MF}). For simplicity, we will assume that $n$ is large, and therefore $\sin^n\theta\ll 1$ for $\lambda<1$.

The single-particle density matrix in the MF ground state is
\begin{equation}\label{eqn: rhoi}
\rho_i = \frac{1}{2} \left( \begin{array}{cc} 
1+\braket{\hs^z} & \braket{\hs^x+i\hs^y} \\ \braket{\hs^x-i\hs^y} & 1-\braket{\hs^z}
\end{array}\right) 
 = \frac{1}{2} \left( \begin{array}{cc} 1 & \sqrt{1-m^2} \\ \sqrt{1-m^2} & 1 \end{array}\right).
\end{equation}
Similarly, the two-particle density matrix can be calculated to be
\begin{equation}\label{eqn: rhoij}
\rho_{ij} = \frac{1}{4} \left( \begin{array}{cccc} 
1+m^2 & \sqrt{1-m^2} & \sqrt{1-m^2} & 1-m^2 \\
\sqrt{1-m^2} & 1-m^2 & 1-m^2 & \sqrt{1-m^2} \\
\sqrt{1-m^2} & 1-m^2 & 1-m^2 & \sqrt{1-m^2} \\
1-m^2 & \sqrt{1-m^2} & \sqrt{1-m^2} & 1+m^2
\end{array}\right).
\end{equation}

Using Eq.~(\ref{eqn: Iij}), the MI between two nodes is
\begin{equation}\label{eqn: IMF result}
\I_{\rm MF} = \frac{1}{2} \left( 1 - \sqrt{1-m^2}\log_2 \frac{1+\sqrt{1-m^2}}{1-\sqrt{1-m^2}} + \frac{2-m^2}{2}\log_2 \frac{2-m^2}{m^2} \right).
\end{equation}
In this case, the MI network is a fully connected weighted network where all links have equal weights. The values of the network measures are $k_i/(n-1)=C_i = \I_{\rm MF}$ and $d_{ij}=1/\I_{\rm MF}$ for $n\gg1$.

The above analysis assumed for simplicity that all spins have the same $Z$. The systems we consider do not satisfy this condition. Then, our MF ansatz is modified to $\ket{\psi} = (\ket{m_1m_2\cdots} + \ket{-m_1-m_2\cdots} )/\sqrt{2}$ (up to a normalization constant). The self-consistent equation for $m_i$ in this case is
\begin{equation} \label{eqn: MF modified}
\frac{m_i}{\sqrt{1-m_i^2}} = \frac{J}{h}\sum_{j\in\mathcal{N}(i)} m_j,
\end{equation}
where $\mathcal{N}(i)$ is the set of neighbors of $i$. The solution to Eq.~(\ref{eqn: MF modified}) is obtained from a set of coupled equations,
\begin{equation} \label{eqn: modified soln}
m_i = \left(1 + \left(\frac{J}{h}\sum_{j\in\mathcal{N}(i)}m_j\right)^{-2} \right)^{-1/2},
\end{equation}
which we solve by iterating Eq.~(\ref{eqn: modified soln}) starting from an initial seed for $\{m_i\}$. 
The MF results in Eqs.~(\ref{eqn: rhoi}) and~(\ref{eqn: rhoij}) are modified to
\begin{equation} \begin{array}{rl}
\hspace*{-3cm} \rho_i &= \frac{1}{2} \left( \begin{array}{cc} 1 & \sqrt{1-m_i^2} \\ \sqrt{1-m_i^2} & 1 \end{array}\right), \\
\hspace*{-3cm} \rho_{ij} &= \frac{1}{4} \left( \begin{array}{cccc} 
1+m_im_j & \sqrt{1-m_i^2} & \sqrt{1-m_j^2} & \sqrt{1-m_i^2}\sqrt{1-m_j^2} \\
\sqrt{1-m_i^2} & 1-m_im_j & \sqrt{1-m_i^2}\sqrt{1-m_j^2} & \sqrt{1-m_j^2} \\
\sqrt{1-m_j^2} & \sqrt{1-m_i^2}\sqrt{1-m_j^2} & 1-m_im_j & \sqrt{1-m_i^2} \\
\sqrt{1-m_i^2}\sqrt{1-m_j^2} & \sqrt{1-m_j^2} & \sqrt{1-m_i^2} & 1+m_im_j
\end{array}\right).
\end{array}
\end{equation}

We will show that even though our assumption of uniform $Z$ is invalid, the values of $k_i$, $C_i$, and $d_{ij}$ are captured well by the solution in Eq.~(\ref{eqn: IMF result}). We also capture the higher moments of $k_i$, $C_i$, and $d_{ij}$, such as the width of their distributions, using the generalized solution in Eq.~(\ref{eqn: modified soln}).

\section{Emergent Networks in Ground States}\label{sec: MI ground state}

An important, and as yet unanswered, question is whether complex networks emerge in ground states of Hamiltonians defined on an imprinted complex network, and if so, what are the resulting properties? Thus, we first calculate the network measures defined in Sec.~\ref{subsec: measures} for emergent networks arising from the ground states of Hamiltonians on the three imprinted networks we consider in Eq.~(\ref{eqn: H}), ER, WS, and BA. Will the emergent networks be complex themselves? Does imprinted complexity beget emergent complexity?

\begin{figure}[t]
\centering
\includegraphics[width=0.8\columnwidth]{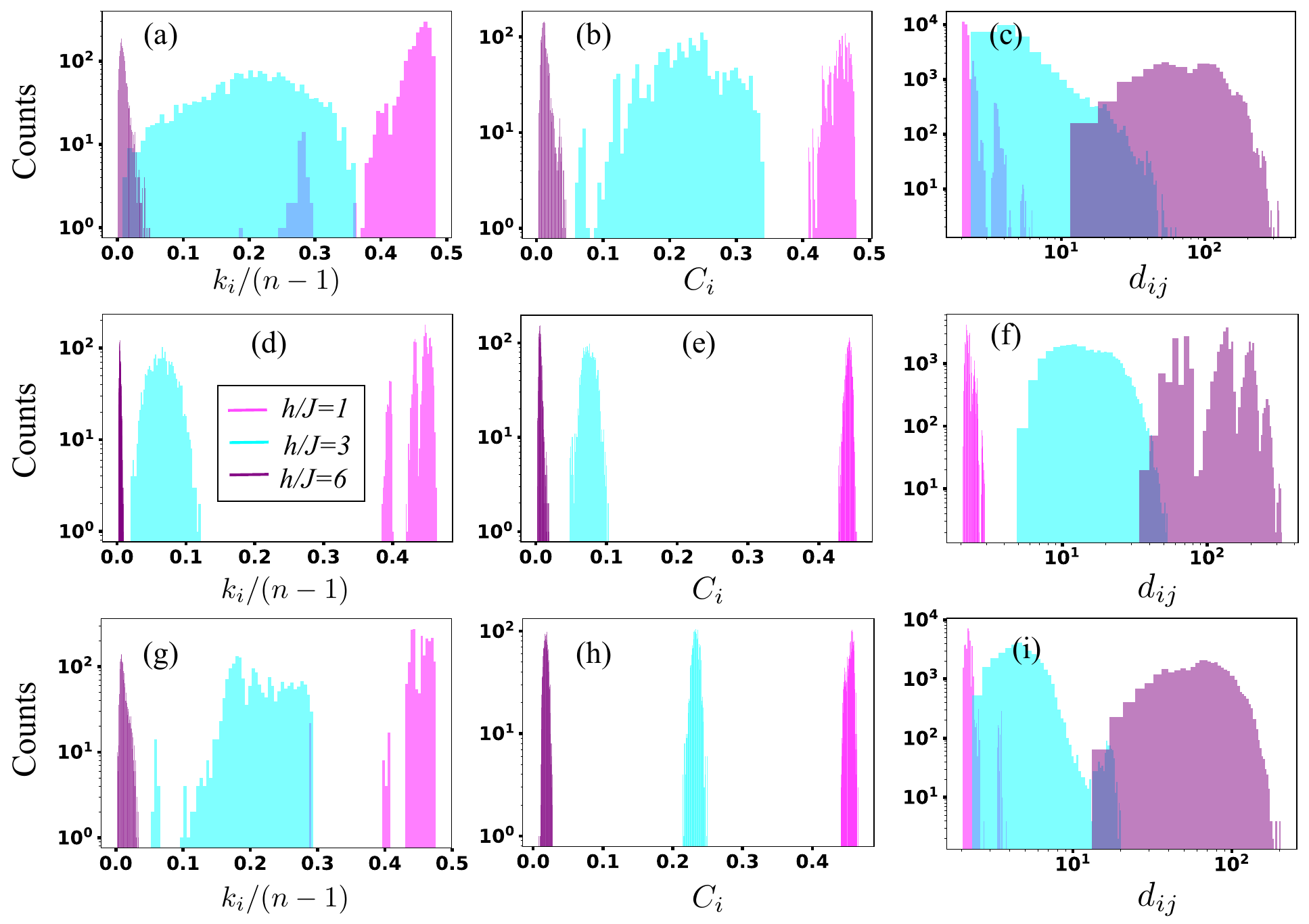}
\caption{\textit{Distributions of complex network measures, the degree $k_i$, clustering $C_i$, and shortest path length $d_{ij}$ in the emergent networks, for an ensemble of imprinted networks generated by random (ER) and complex (WS and BA) network models.} (a) Distributions of $k_i$, (b) $C_i$, and (c) $d_{ij}$ for the three different values of $h/J$ (indicated above the histograms), when the imprinted networks are generated by ER and complex model with $n=20$ and $p=0.26$. (d-f) Same as (a-c) for imprinted WS networks with $K=4$ and $p=0.5$. (g-i) Same as (a-c) for imprinted BA networks with $m=3$. The degree and clustering shift to smaller values, and the path lengths to larger values, as $h/J$ increases. The distributions resemble those of random networks: the degree distributions do not have long tails, and $C_i$ and $k_i/(n-1)$ are of the same order.}
\label{fig3}
\end{figure}

\begin{figure}[t]
\centering
\includegraphics[width=0.8\columnwidth]{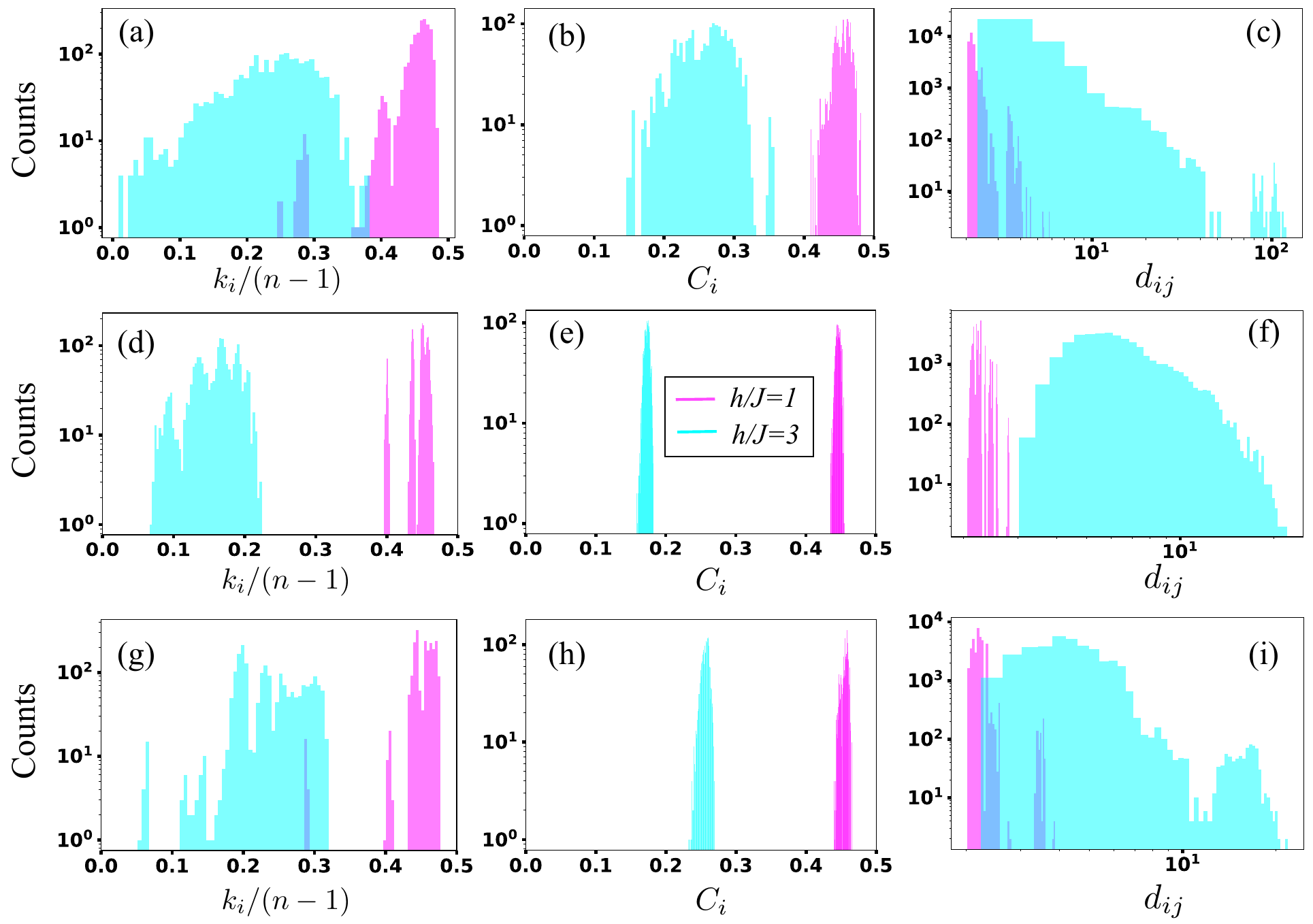}
\caption{\textit{Mean field distributions of the degree $k_i$, clustering $C_i$, and shortest path length $d_{ij}$ in the emergent networks, for an ensemble of imprinted networks generated by random (ER) and complex network (WS and BA) models.} The imprinted networks are generated with the same parameters as Fig.~\ref{fig3}. The mean field distributions qualitatively capture the features obtained from exact diagonalization in Fig.~\ref{fig3}.}
\label{figmf}
\end{figure}

\begin{figure}[b]
\centering
\includegraphics[width=0.95\columnwidth]{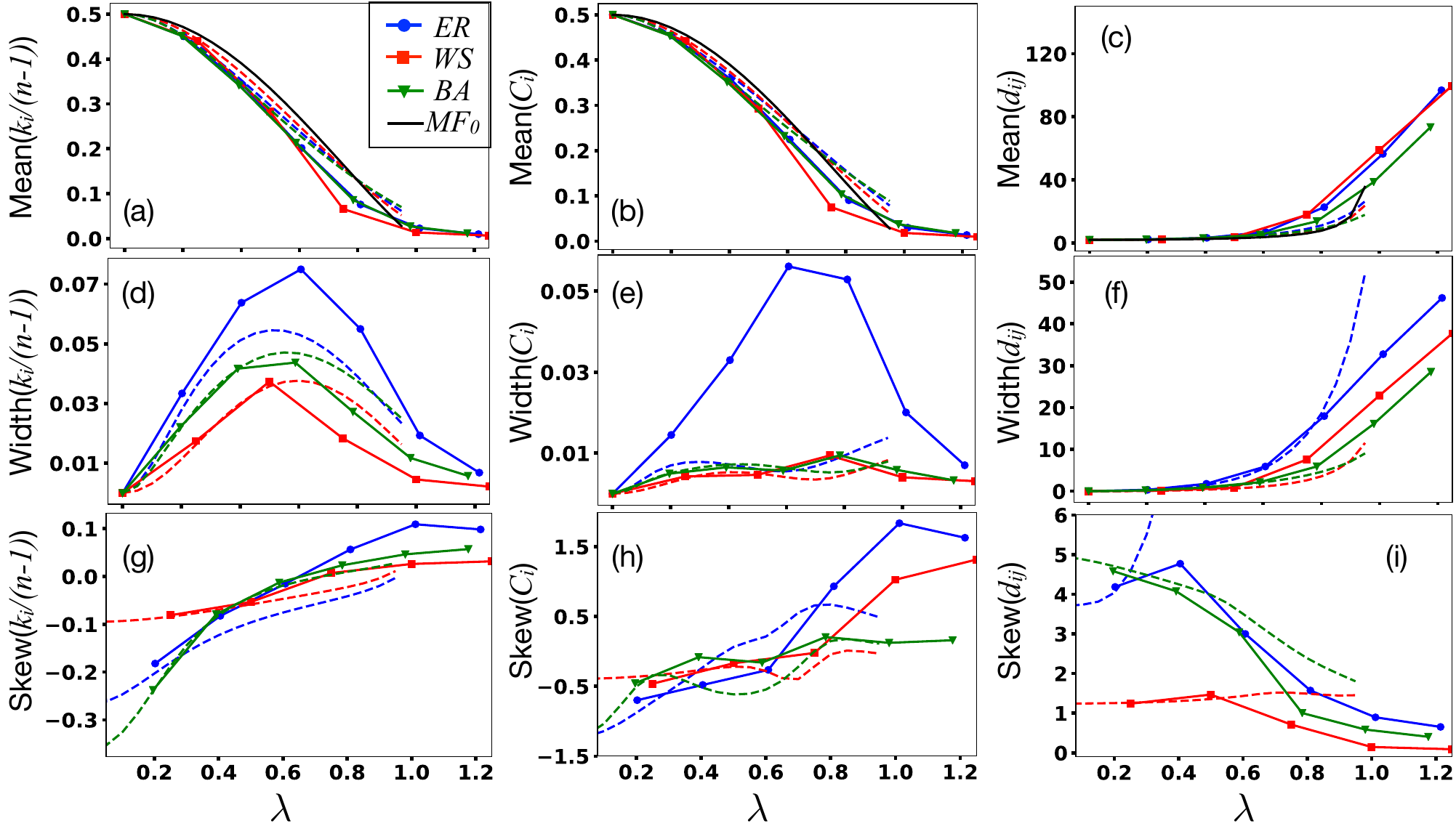}
\caption{\textit{Moments of the distributions of network measures from Fig.~\ref{fig3}.} (a) Mean value of $k_i/(n-1)$ versus $\lambda=h/(ZJ)$ in the emergent networks, for imprinted networks drawn from ER, WS, and BA models with the same parameters as in Fig.~\ref{fig3}. (b) Mean value of $C_i$ versus $\lambda$, and (c) of $d_{ij}$ versus $\lambda$, in the emergent networks. (d-f) plot the width of the distributions of $k_i$, $C_i$, and $d_{ij}$, and (g-i) plot the distributions' skewness. The solid curves plot the result of an exact calculation, the dashed curves plot the generalized mean field prediction obtained from Eq.(~\ref{eqn: modified soln}), and the black curves in (a-c) plot the uniform mean prediction, MF$_0$, from Eq.~(\ref{eqn: IMF result}). The mean field curves accurately capture the mean values of the measures in (a-c), but have differences in the higher moments (d-i).
}
\label{fig4}
\end{figure}

In Figs.~\ref{fig3}-\ref{fig4}, we consider an ensemble of $100$ imprinted networks with $n=20$ generated from all three network models, ER, WS and BA. The ER networks have $p=0.26$ and on average, $Z=(n-1)p=4.94$. The value of $p$ was optimally chosen to yield connected ER networks; $p<0.26$ yields mostly disconnected ER networks for $n=20$. The WS networks have $K=4, p=0.5$ and $Z=K=4$. The BA networks have $m=3$ and $Z=2m(n-m)/n=5.1$. We note that due to finite $n$ and since $K$ and $m$ are integers, we are restricted to consider networks with different $Z$. However, this is not a limitation, since our goal is not comparing the network measures for different ensembles of imprinted networks with different structures and different $Z$. Instead, our goals are quantifying the complexity of the emergent networks, and investigating their response to network attacks. To mitigate the effects of different $Z$, we will plot the moments of the distributions of network measures versus $\lambda = h/(ZJ)$.

Figure~\ref{fig3} plots the distribution of $k_i/(n-1)$, $C_i$ and $d_{ij}$ obtained from an exact calculation for ER networks in (a-c), for WS networks in (d-f), and BA networks in (g-i). The purple, blue and pink histograms show the distributions for normalized transverse field $h/J=1, 3$ and $6$ respectively. The transverse field drives a quantum phase transition in the much simpler transverse Ising model and has been shown to give rise to complexity in the emergent network even for a simple nearest-neighbor spin chain near the quantum critical point~\cite{carr2010understanding,valdez2017quantifying,sundar2018complex}.

A few features stand out. The distribution of $k_i\ (C_i)$ is a Dirac delta function at $k= (n-1)/2\ (C=1/2)$ for $h=0$ (not shown), and the distribution of $d_{ij}$ is a Dirac delta at $d=2$. This is because at $h=0$, the ground state is the GHZ state, where all the spins point along $\pm\hz$. As $h$ is increased, the distributions of $k_i$ and $C_i$ get wider first and then narrower, and the centre shifts towards $0$. The distribution of $d_{ij}$ gets wider and shifts towards larger values. At $h/J=\infty$ (not shown), all spins point along $\hx$, $k_i=0$, $C_i=0$ and $d_{ij}=\infty$.

The distribution of $k_i$ for the emergent networks does not have a long tail or a power law, and is strikingly different from the distributions for the imprinted networks in Fig.~\ref{fig1}. Moreover, the distribution of $C_i$ is of the same order as $k_i/(n-1)$, which is typical for random networks. Therefore, we arrive at the important conclusion from Fig.~\ref{fig3} that the emergent MI networks do not show complexity. 

Figure~\ref{figmf} plots the MF distribution of $k_i/(n-1)$, $C_i$ and $d_{ij}$ obtained by applying Eq.~(\ref{eqn: modified soln}) for an ensemble of ER networks in (a-c), WS networks in (d-f), and BA networks in (g-i). The blue and pink histograms show the distributions for normalized transverse field $h/J=3$ and $6$ respectively, the same color scheme as Fig.~\ref{fig3}. Comparing the two figures, the MF distributions are qualitatively similar to the distributions obtained from the exact calculation in Fig.~\ref{fig3}. Thus, we observe that MF theory is qualitatively adequate to capture the emergent network features.

Figure~\ref{fig4} plots the moments of the distributions of $k_i/(n-1)$, $C_i$ and $d_{ij}$. It confirms our claim above that $C_i$ is of the same order as $k_i/(n-1)$. Moreover, we also find that the curves overlap each other when plotted as a function of $\lambda$, although the networks have different $Z$. The solid blue/red/green curves are obtained numerically from exact diagonalization, while the dashed curves are obtained from the MF theory used in Eq.~(\ref{eqn: modified soln}) and Fig.~\ref{figmf}. The mean values (panels (a-c)) in MF theory and the exact calculation agree well, while the disagreement in the higher moments is larger.

The solid black curve in Figs.~\ref{fig4}(a-c) is the MF prediction for the moments for $k_i/(n-1)$, $C_i$, and $d_{ij}$, assuming a uniform degree for all the nodes [Eq.~(\ref{eqn: IMF result})], and captures the trend in the exact calculations. For simplicity, we will use Eq.~(\ref{eqn: IMF result}) in the rest of the paper to calculate the MF values for the average $k_i/(n-1)$, $C_i$, and $d_{ij}$, and we will denote this MF$_0$. We note that MF$_0$ correctly captures the mean values, while the more general MF captures the higher moments too.

\section{Attacks unique to quantum networks: Decoherence as modeled by projective spin measurement}\label{sec: attacks}

\begin{figure}[t]
\centering
\includegraphics[width=0.8\columnwidth]{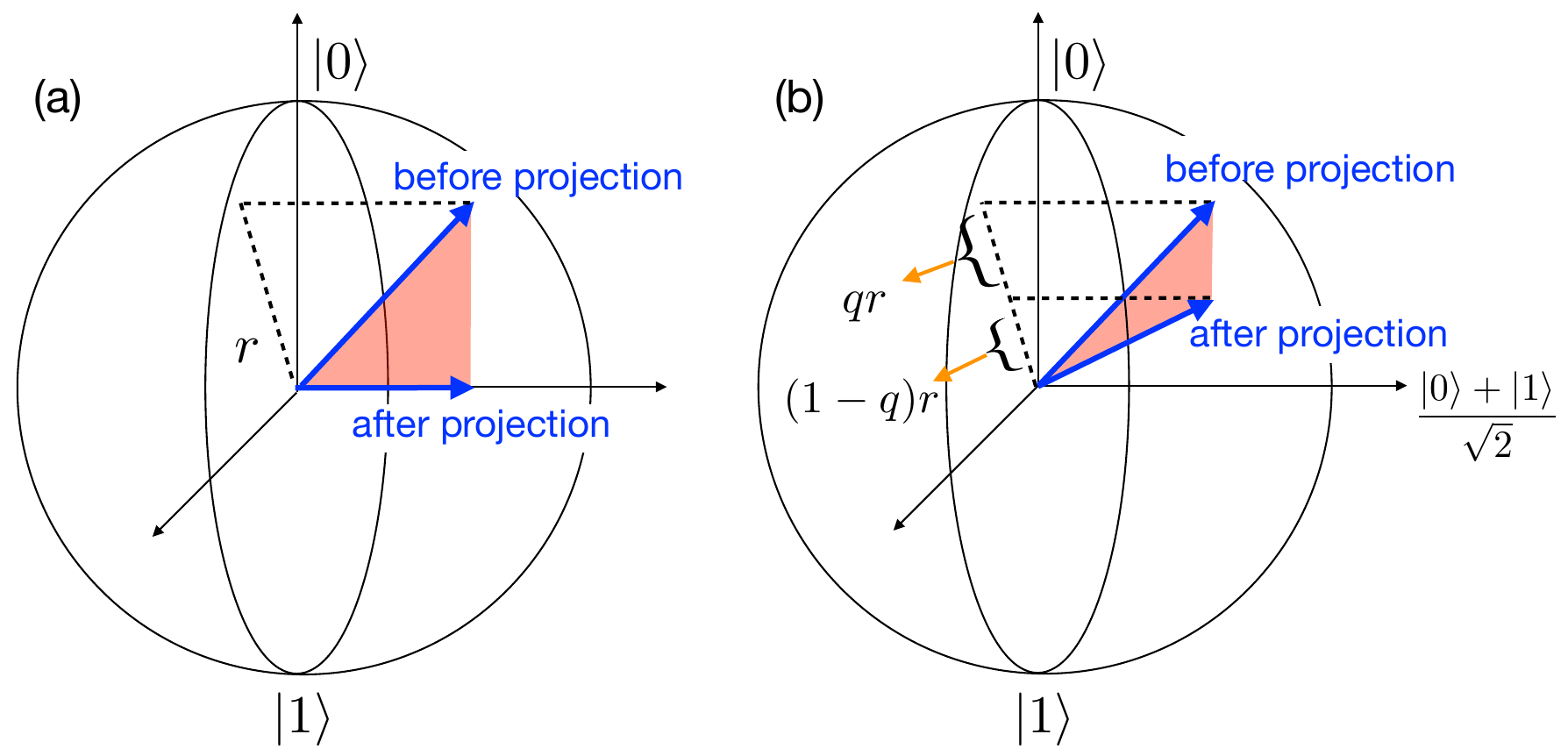}
\caption{\textit{Projective measurement attack on a spin}. (a) A complete projective measurement projects the spin's Bloch vector onto the axis of measurement (here, $\hx$), reducing its component on the plane orthogonal to measurement (here, the $y$--$z$ plane), from an initial value of $r$ to a final value of $0$. (b) A partial projective measurement reduces its component on this plane to $(1-q)r$.}
\label{fig: bloch sphere X}
\end{figure}

Network scientists consider various kinds of attacks on networks. Common kinds of attacks are adding or removing nodes or links, where the nodes or links are chosen randomly or preferentially based on the degree or centrality measures of the attacked node. They find that the response of the network depends on the kind of network -- complex or random -- and how the attacked nodes are chosen -- random or preferential~\cite{holme2002attack, nie2015new, gallos2006attack, nguyen2019conditional, chaoqi2018multi, vsimon2017combined}. We briefly study the response of the imprinted networks to classical attacks in~\ref{sec: appendix A}.

Quantum networks allow a richer set of attacks than attacks on classical networks. This is essentially because quantum systems have more degrees of freedom. In the systems we consider, each node has a spin whose local Hilbert space dimension is $2$. The initial state is a pure state in a Hilbert space with $2^n$ dimensions, the usual exponentially growing state space of quantum many-body systems, described by a wave function $\ket{\psi}$. The attacks we consider here project the pure state wave function into a mixed state, which can be described by a density matrix $\rho$. For a pure state, $\rho = \ket{\psi}\bra{\psi}$.

Specifically, we consider projective measurement of spins. This attack does not have a classical counterpart. A projective measurement is a probabilistic process which collapses a spin onto the direction along which it is measured, and is represented on the Bloch sphere as projecting the spin's Bloch vector onto the axis of measurement. Projection can happen along any direction. Figure~\ref{fig: bloch sphere X}(a) illustrates projection onto $\hx$. In the density matrix picture, when $\rho$ for a single spin is written in the basis set by the direction of measurement (which is $\hx$ here), a complete projective measurements sets its off-diagonal elements to zero,
$$ \rho = \left( \begin{array}{cc} \rho_{11} & \rho_{12} \\ \rho_{21} & \rho_{22} \end{array}\right) \rightarrow \left( \begin{array}{cc} \rho_{11} & 0 \\ 0 & \rho_{22} \end{array}\right).$$
We consider a generalized version where the projective measurement only partially suppresses the off-diagonal elements,
$$ \rho = \left( \begin{array}{cc} \rho_{11} & \rho_{12} \\ \rho_{21} & \rho_{22} \end{array}\right) \rightarrow \left( \begin{array}{cc} \rho_{11} & (1-q)\rho_{12} \\ (1-q)\rho_{21} & \rho_{22} \end{array}\right),$$
where we denote $q$ as the strength of the projective measurement. This is illustrated in Fig.~\ref{fig: bloch sphere X}(b).

\begin{figure}[t]
\centering
\includegraphics[width=0.95\columnwidth]{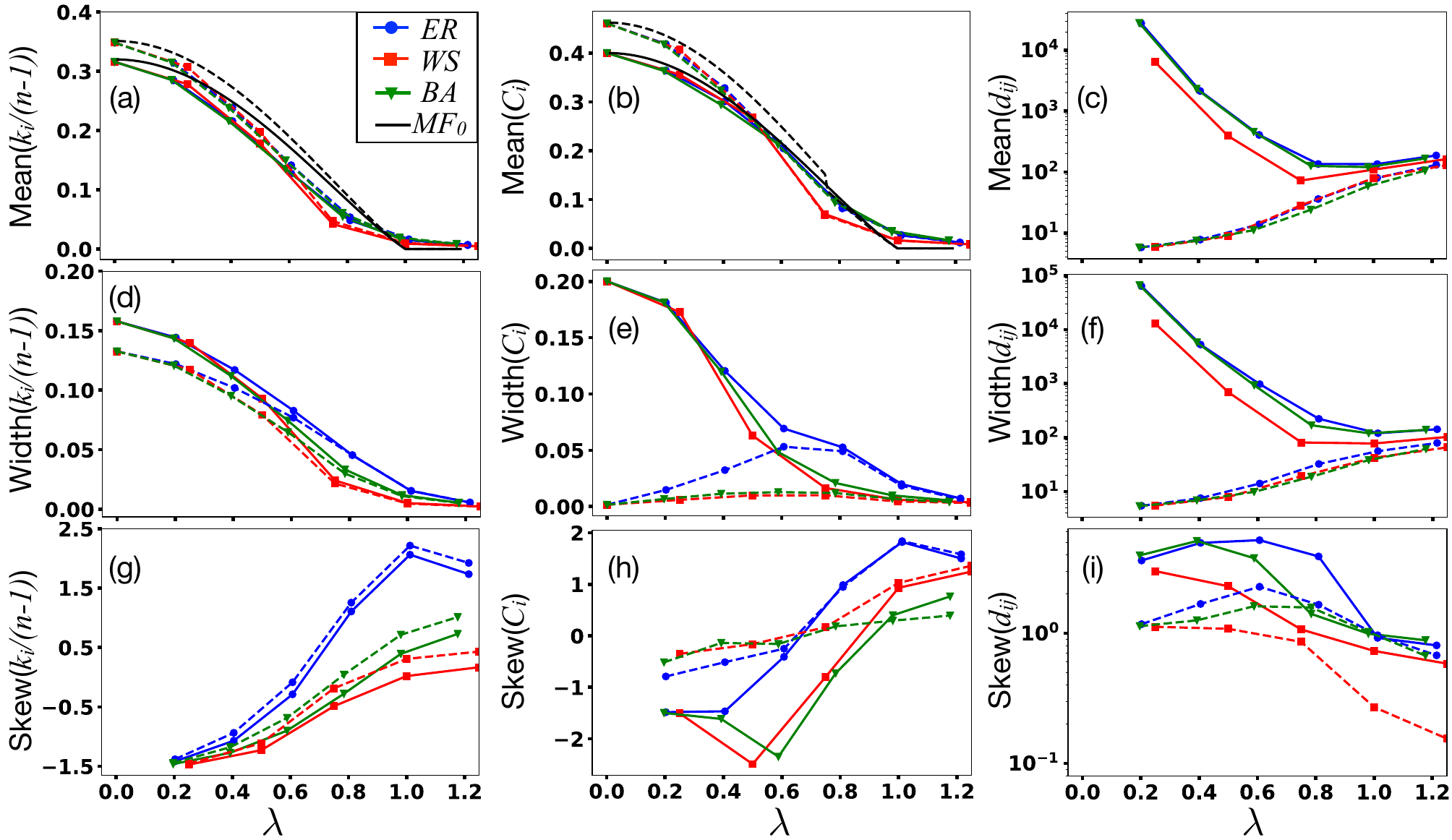}
\caption{\textit{Moments of the distributions of the measures $k_i$, $C_i$, and $d_{ij}$ in the emergent network, after $20\%$ of the nodes in the ground state of each imprinted network are randomly picked and projectively measured along $\hx$.} (a-c) Solid curves plot the mean values of the measures for complete projective measurements ($q=1$), and dashed curves plot the same for partial projective measurements ($q=0.5$). Black curves plot the uniform mean field (MF$_0$) prediction. (d-f) plot the width of the distributions of the measures, and (g-i) plot the distributions' skewness. The imprinted networks used here are the same as those in Fig.~\ref{fig3}. The moments of the measures are similar for both measurement strengths, and the mean values are captured well by MF$_0$.
}
\label{fig6}
\end{figure}

\begin{figure}[t]
\centering
\includegraphics[width=0.95\columnwidth]{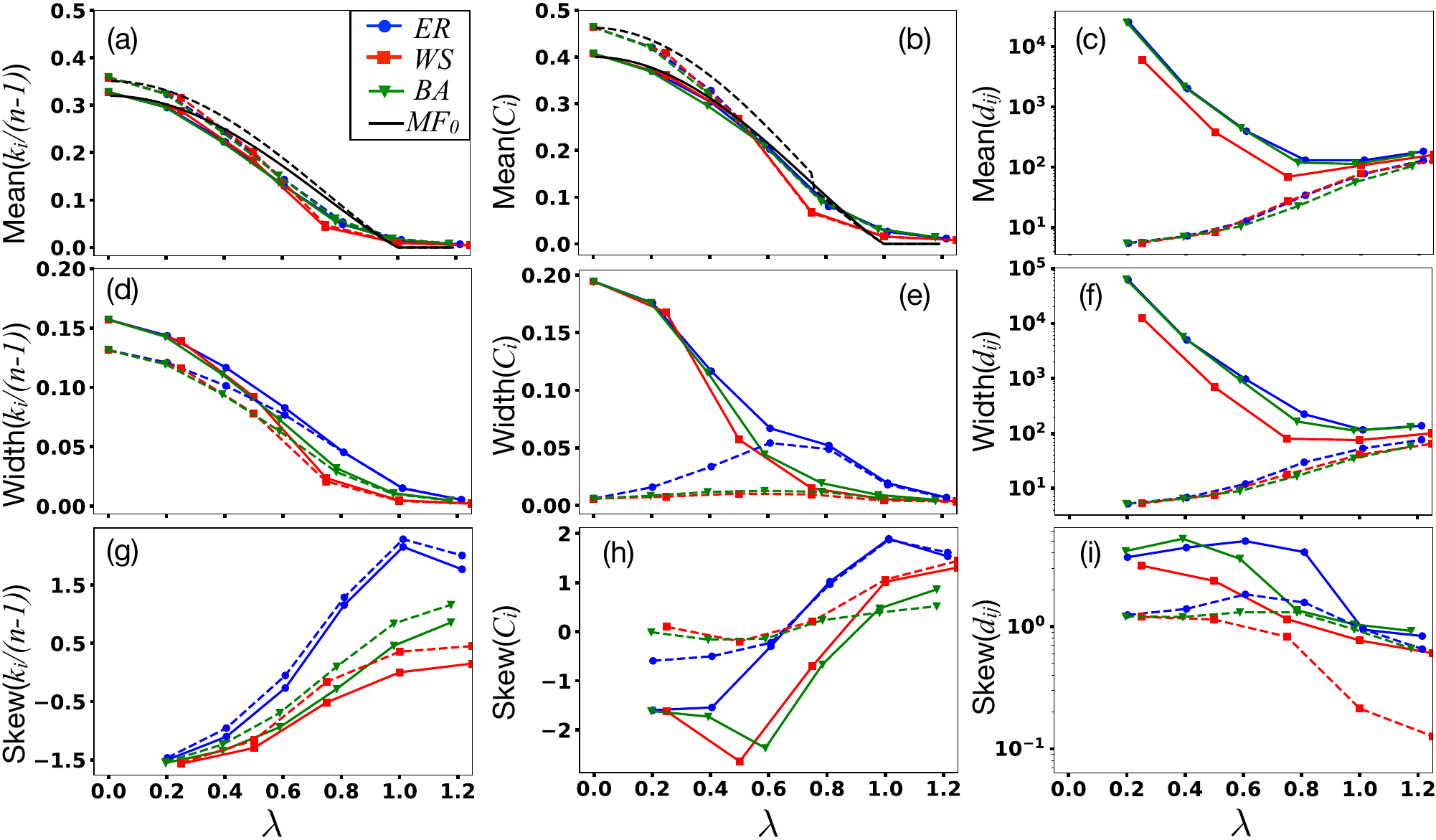}
\caption{ \textit{Moments of the distributions of the measures $k_i$, $C_i$, and $d_{ij}$ in the emergent network, after $20\%$ of the nodes in the ground state of each imprinted network are preferentially picked and projected onto $\hx$.} The different panels plot the same quantities as Fig.~\ref{fig5} with the same color coding. There is almost no difference between the results for targeted attacks (this figure) and random attacks [Fig.~\ref{fig6}].
}
\label{fig8}
\end{figure}

\section{Response to attacks}\label{sec: results}

We choose a fraction of the nodes in the network. These nodes will be either picked randomly, or preferentially by the degree of $\I_{ij}$. In Sec.~\ref{subsec: projective}, we then projectively measure the spins on these nodes, with a strength $q$ along $\hx$. This is the main focus of our work. We discuss projections along $\hz$ in~\ref{sec: appendix B}. In Sec.~\ref{subsec: other}, we also briefly discuss other possible attacks that may occur on quantum complex networks, which may peak the reader's interest to perform future robustness studies.

\subsection{Projective attacks}
\label{subsec: projective}

We consider networks with $n=20$ spins, and projectively measure $20\%$ of the spins. We vary the measurement strength $q$, and the strategy used to choose the projected spins. We perform these attacks on the ground states of an ensemble of ER, WS, and BA networks, and calculate the complex network of the emergent MI network in all three cases. In addition to averaging over the ensemble of $100$ networks for each network model, we also average over $50$ realizations of the attack, i.e. $50$ different combinations of the spins that we project, for the ground state on each imprinted network. 

In Fig.~\ref{fig6}, we consider the case that the networks are attacked by a projective measurement along $\hx$, where the attacked spins are chosen randomly. Figure~\ref{fig6} plots the moments of the distributions of $k_i/(n-1)$, $C_i$, and $d_{ij}$ after this attack. As in Fig.~\ref{fig4}, panels (a-c) plot the average of $k_i/(n-1)$, $C_i$, and $d_{ij}$, panels (d-f) plot the width of their distributions, and panels (g-i) plot the skew. Panels (a-c) have eight curves each. The solid blue/red/green curves correspond to the case that the attacked nodes in each ER/WS/BA network are completely projected ($q=1$) to the $\hx$ direction. The dotted curves correspond to partially projecting the attacked spins to $\hx$ with strength $q=0.5$.

Remarkably, we find that the three solid curves for exact results in Figs.~\ref{fig6}(a-c) again overlap with each other when plotted versus $\lambda$, and the three dotted curves also overlap with each other. This occurs even though the imprinted networks have very different structures and different $Z$. Our calculations indicate that imprinted complex networks (such as WS and BA networks) do not show a more robust response to projective measurement along $\hx$, as compared to non-complex (i.e ER) networks. This is completely different from the case of classical network attacks, where the response of the system is highly dependent on whether the network is complex.

We also compare the exact results obtained from exact diagonalization with predictions from MF$_0$, shown as solid and dashed black curves in Figs.~\ref{fig6}(a-b). When a node $i$ is attacked with a projective measurement along $\hx$, its magnetization $\braket{\hs^z_i}$ and all correlations $\braket{\hs^z_i\hs^z_j}$ get shrunk by $(1-q)$. Since $\braket{\hs^z_i} = 0$ due to $Z_2$ symmetry, the node's single-particle density matrix after projection is left unchanged. The two-particle density matrix $\rho_{ij}$ after an attack on node $i$ becomes
\begin{equation}\label{eqn: attackX}
\hspace*{-2.5cm}
\rho_{ij} = \frac{1}{4} \left( \begin{array}{cccc} 
1+m^2(1-q) & \sqrt{1-m^2} & \sqrt{1-m^2} & (1-q)(1-m^2) \\
\sqrt{1-m^2} & 1-m^2(1-q) & (1-m^2) & \sqrt{1-m^2} \\
\sqrt{1-m^2} & (1-m^2) & 1-m^2(1-q) & \sqrt{1-m^2} \\
(1-q)(1-m^2) & \sqrt{1-m^2} & \sqrt{1-m^2} & 1+m^2(1-q)
\end{array}\right).
\end{equation}
If both nodes $i$ and $j$ are attacked, then $\rho_{ij}$ has a similar form to Eq.~(\ref{eqn: attackX}), with $(1-q)$ replaced by $(1-q)^2$. The MI can then be calculated using Eqs.~(\ref{eqn: Iij}) and~(\ref{eqn: attackX}).

The black curves in Figs.~\ref{fig6}(a-c) plot the MF$_0$ prediction for the mean value of $k_i/(n-1)$, $C_i$, and $d_{ij}$. These quantitatively differ from the exact results, but are qualitatively similar. MF$_0$ does not correctly predict any higher moments of the distributions.

In Fig.~\ref{fig8}, we pick the nodes to be attacked preferentially, with the probability of a node $i$ to be attacked given by $k_i/\sum_j k_j$. The attacks are again projective measurements along $\hx$ with $q=0.5$ (dotted) or $q=1$ (solid). We find almost no difference to the case of random attacks. This is a surprising finding, since the response of complex networks to classical network attacks is known to depend on whether the attacked nodes are picked randomly or preferentially.

Our calculations highlight two striking aspects. First, they indicate that (imprinted) complex networks respond similarly to projective measurement attacks as non-complex networks, and respond similarly when the attacks are random or preferential. This is evident from the overlapping curves for the three imprinted networks in Figs.~\ref{fig6}-\ref{fig8}, and is completely different from the case of classical network attacks -- network attacks with a classical counterpart, such as deleting nodes -- where the response to the attacks depend heavily on the structure of the imprinted networks. The main reason for this is that the attacks are made on the emergent networks, and these networks do not seem to be complex. Moreover, MF$_0$ as well as the general MF theory qualitatively capture the mean values of $k_i/(n-1)$, $C_i$, and $d_{ij}$, although MF$_0$ does not capture any higher moments, and the general MF theory has quantitative disagreements for all higher moments of the distributions. The accuracy of MF and MF$_0$ is likely due to the large degree of nodes in the imprinted networks, consistent with its accuracy in regular lattices with a large $Z$ in condensed matter systems. As we show later, the results might deviate from MF theory when the degree is made smaller.

\begin{figure}[t]\centering
\includegraphics[width=1.0\columnwidth]{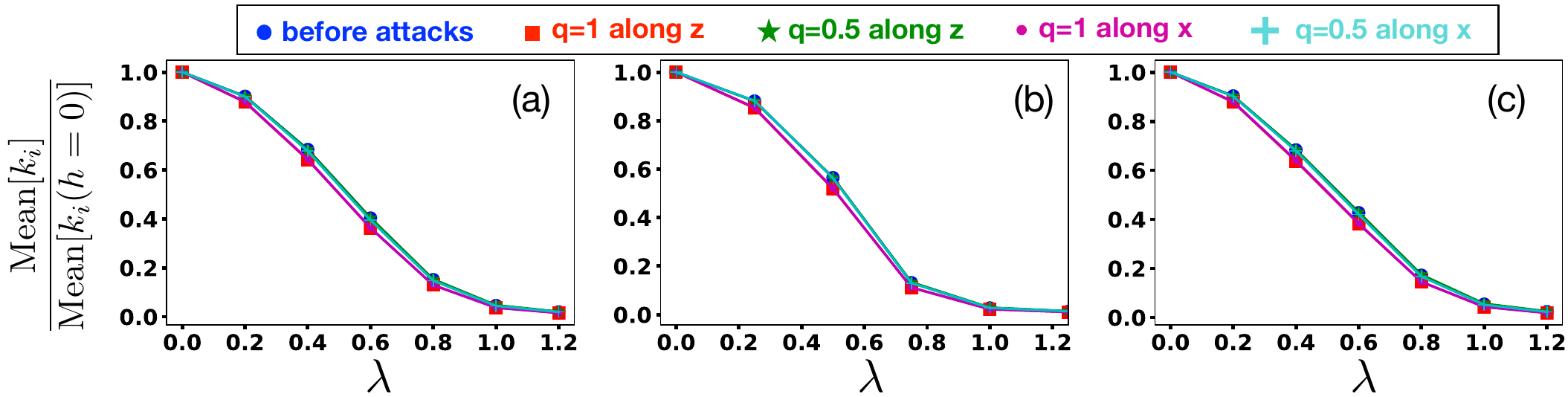}
\caption{\textit{Collapse of Mean[$k_i$]/Mean[$k_i(h=0)$], before and after projective measurement attacks.} The three panels consider imprinted ER, WS, and BA networks respectively. Each panel has five (nearly overlapping) curves for Mean[$k_i$]/Mean[$k_i(h=0)$] versus $\lambda=h/(ZJ)$, for the case before attacks, and for the cases of random or preferential projective measurements along $\hz$ or $\hx$ with $q=1$.
}
\label{fig: before vs after}
\end{figure}

\begin{figure}[t]
\centering
\includegraphics[width=0.95\columnwidth]{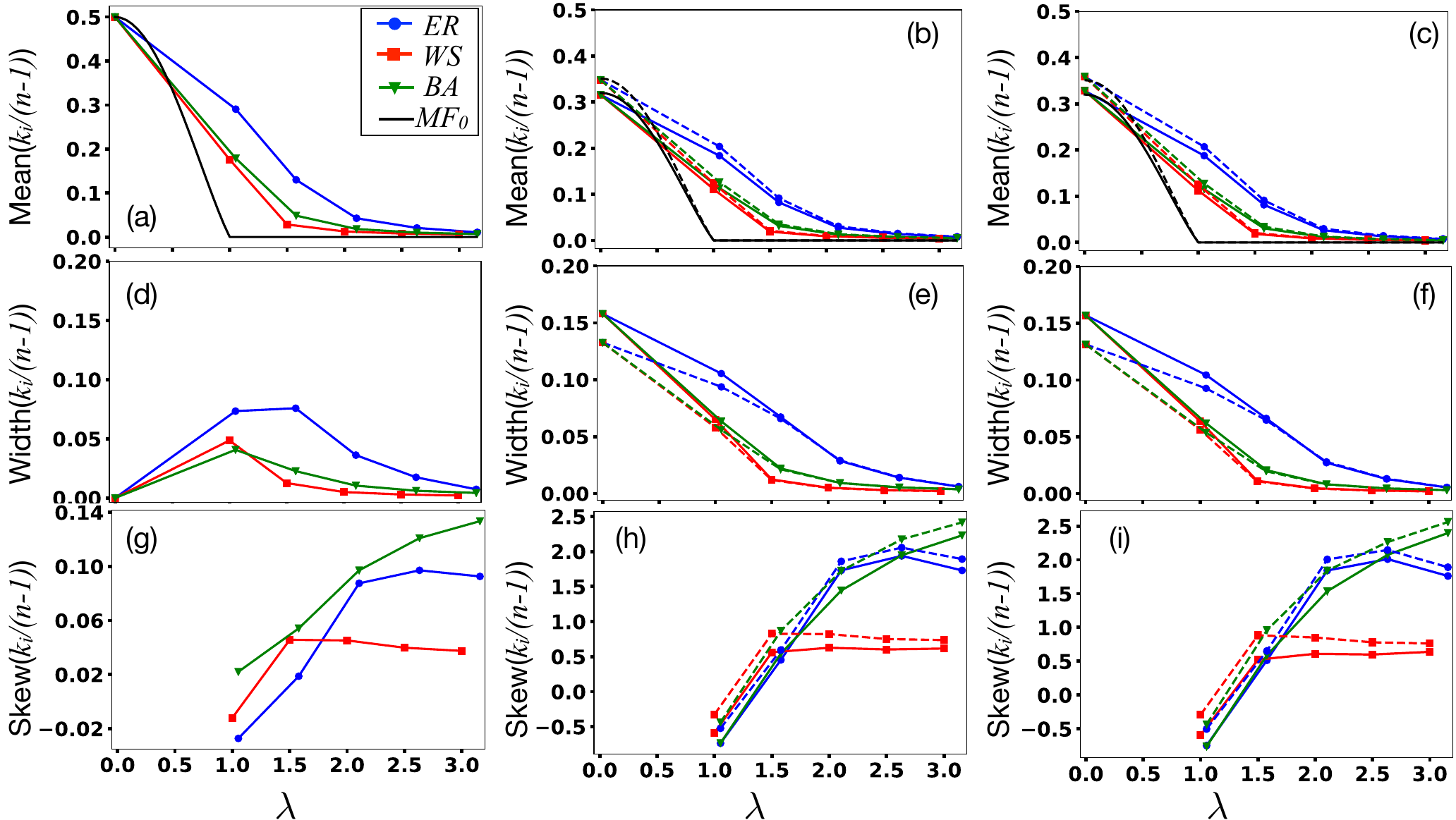}
\caption{ \textit{Moments of the distributions of $k_i/(n-1)$ versus $\lambda=h/(ZJ)$, before and after projective attacks.} We consider imprinted ER networks with $p=0.1$, WS networks with $K=2$ and $p=0.5$, and BA networks with $m=1$.  Panels (a,d,g) consider networks before projective attacks, (b,e,h) consider $20\%$ of the nodes in the ground state of each imprinted network to be randomly picked and projected onto $\hx$, and (c,f,i) pick the projected nodes preferentially. The color coding is the same as in Fig.~\ref{fig5}. The mean degree in the imprinted networks is smaller than those in Figs.~\ref{fig4} and~\ref{fig5} and outside the validity of MF theory, therefore the distributions do not agree with MF theory (black curves).
}
\label{fig9}
\end{figure}

\begin{figure}[t]
\centering
\includegraphics[width=0.95\columnwidth]{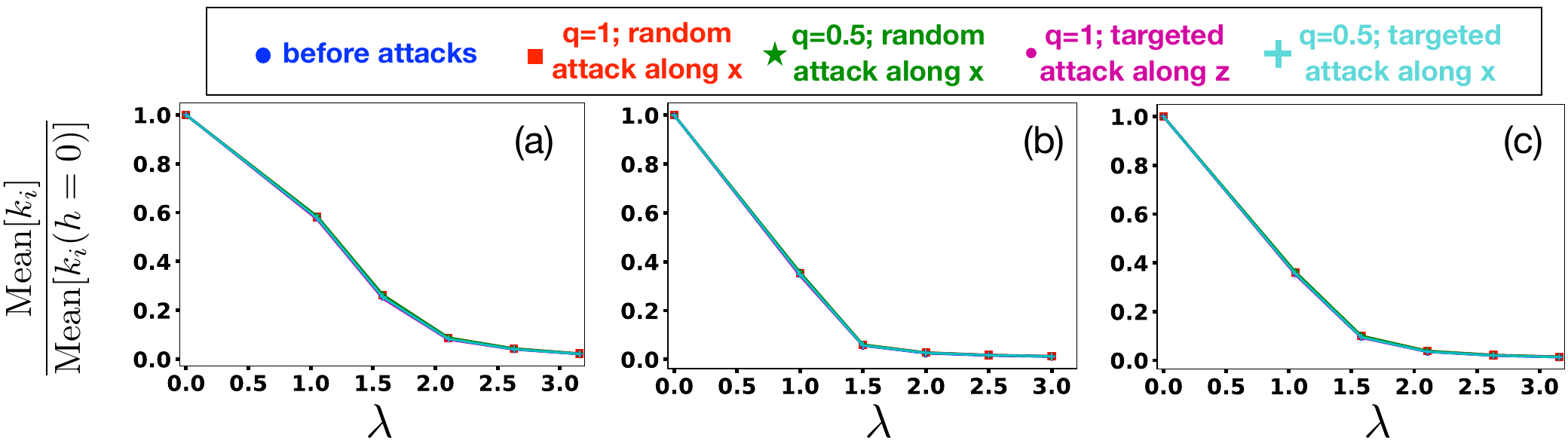}
\caption{\textit{Collapse of Mean[$k_i$]/Mean[$k_i(h=0)$], before and after projective measurement attacks.} The three panels respectively consider the imprinted ER, WS, and BA networks in Fig.~\ref{fig9}. Each panel has five (nearly overlapping) curves for Mean[$k_i$]/Mean[$k_i(h=0)$] versus $\lambda=h/(ZJ)$, for the case before attacks, and for the cases of random or preferential projective measurements along $\hx$ with $q=1$ or $q=0.5$.
}
\label{fig10: before vs after sparse}
\end{figure}

The second striking aspect of our calculation, illustrated in Fig.~\ref{fig: before vs after}, is that we also observe a collapse of the curves for average $k_i$ when it is normalized by the average $k_i(h=0)$, for the cases before attacks and after attacks (for all the four kinds of attacks considered in Figs.~\ref{fig6}-\ref{fig7} and~\ref{fig5}-\ref{fig8}). This indicates that projective measurement attacks only rescale the average value of $k_i$ before the attack by a constant factor, irrespective of the Hamiltonian parameters or the imprinted network's structure.

Finally, we consider an ensemble of imprinted networks that are expected to lie outside the validity of MF theory. To this end, we consider ER networks with $p=0.1$, WS networks with $K=2$ and $p=0.5$, and BA networks with $m=1$, all of which have a smaller $Z$ than before. We only consider connected networks, and discard disconnected networks. Figure~\ref{fig9}(a,d,g) plots the moments of the density $k_i/(n-1)$ for these networks. We clearly see that the mean density is not captured by MF$_0$. Unlike the case of dense imprinted networks, we also see here a difference between the mean value of $k_i$ in the three classes of networks.

Performing projective attacks on these networks, we find that MF$_0$ theory again does not capture the mean density. Figure~\ref{fig9}(b,e,h) plot the moments of the density $k_i/(n-1)$ after $20\%$ of the nodes are randomly picked and projected along $\hx$., and Fig.~\ref{fig9}(c,f,i) plot the moments of $k_i/(n-1)$ after $20\%$ of the nodes are preferentially picked and projected along $\hx$.

Nevertheless, as shown in Fig.~\ref{fig10: before vs after sparse}, we again find that when the average $k_i$ is rescaled by the average $k_i(h=0)$, the curves before and after projective attacks collapse onto each other. This indicates that even when MF does not capture the distributions of $k_i$ (and possibly other network measures too), the dominant effect of projective attacks on the average measure is only a simple rescaling.

The small network size that we consider is a potential concern in comparing the robustness of complex and random networks to random versus targeted attacks. We are limited to $n=20$ for calculating the ground state with exact diagonalization. Larger networks can be studied using more sophisticated numerical techniques, or realized in experiment. Network size is, however, much less of a concern for calculating measures of classical networks. Exploiting this, we investigate the robustness of complex and random networks to classical attacks at $n=20$ and $n=54$ in~\ref{sec: appendix A}, as well as consider a larger ensemble of classical networks. For the classical attack, we delete nodes, since this resembles projectively measuring a node with $q=1$, which sets the MI between the measured node and all other nodes as $0$. We find that BA networks with $n=54$ are more robust to random deletions than targeted deletions, as expected, and this robustness is less prominent for networks with $n=20$. An experimental realization of the quantum networks considered in this paper will conclusively demonstrate whether the robustness to random classical attack, i.e. node deletions, is also extended to projective measurement attacks.

\subsection{Non-projective quantum network attacks}\label{subsec: other}

Quantum networks are subject to many kinds of attacks. In the following, we briefly sketch some forms of attack for future study.

Decoherence effects can be modeled by a quantum trajectories approach or by a master equation, for example the Lindblad master equation. There are many sources of possible decoherence. For the master equation approach, each form of decoherence is identified by Lindblad operators. We list a couple below.

Single-qubit dephasing along $\hx$ is modeled by the master equation
\begin{equation}
\partial_t \rho = -i[\hH,\rho] + \gamma\sum_i \hs^x_i\rho\hs^x_i - \rho
\end{equation}
Collective dephasing is modeled by the master equation
\begin{equation}
\partial_t \rho = -i[\hH,\rho] + \gamma \left( \hs^x_{\rm tot}\rho\hs^x_{\rm tot} - \frac{1}{2}(\hs^x_{\rm tot})^2\rho - \frac{1}{2}\rho(\hs^x_{\rm tot})^2 \right).
\end{equation}
Both these forms of dephasing suppress density matrix elements that are off-diagonal in the $\hx$ basis. We expect these to have the same effect as the projective measurements we consider, which also suppress off-diagonal elements.

Depolarizing is modeled by the master equation
\begin{equation}
\partial_t \rho = -i[\hH,\rho] + \gamma \left(2^{-n} - \rho \right).
\end{equation}
The solution to this equation is $\rho(t) = e^{-\gamma t}e^{-i\hH t}\rho e^{i\hH t} + (1-e^{-\gamma t})/2^n$. The two-particle density matrices, which are used to define $\I_{ij}$, also have the same form. Insofar as MF theory qualitatively captures the mean value of $k_i/(n-1)$, $C_i$, and $d_{ij}$ before the attack, we expect it to capture them after depolarizing attack as well.

Recently, it has been shown for the transverse Ising model that a \emph{multichannel} Lindblad approach is needed to correctly predict equilibration times, which provide a bound on decoherence times~\cite{jaschke2019}. Thus attacks due to a reservoir may address global eigenstates, i.e., the entire wave function -- not just local qubits. We expect the same issues to arise for quantum complex networks based on spins with an Ising coupling.

Other attack scenarios include the usual removal of nodes one finds in classical complex network robustness studies, probes of reduced Hilbert spaces on a subset of qubits, and using a single immersed qubit as a probe of complex network structure~\cite{nokkala2016complex}.

\section{Conclusions}\label{sec: summary}
We used a network-science approach to study ground states of spin models on complex networks, and the effects of decoherence on the ground states. The main questions we asked were (1) whether complexity in the Hamiltonian (the imprinted network) leads to complexity in the emergent network defined in the state, and (2) what are the consequences of this emergence to the robustness of the state to decoherence. We quantified the complexity of the state's emergent mutual information network, which is a weighted network with link weights given by the pairwise mutual information between spins, by calculating complex network measures for this network. We simulated decoherence by performing a series of projective measurement attacks on the networks.

Answering the first question, we found that emergent networks differ significantly from the imprinted networks, and do not show signs of complexity, even if the imprinted networks are complex. The curves of average values of the emergent networks' measures versus a dimensionless parameter nearly overlapped with each other for different imprinted networks with a large average degree ($Z\gtrsim4$), in the ground state before the attacks, and in the state after the attacks. We find that mean field theory accurately captures the average values of the emergent network measures, for all three classes of imprinted networks, but had quantitative differences with the exact results in the distributions' higher moments. Thus, while the mean of the distributions is mean-field, the width, skew, and kurtosis show quantum many-body effects. Importantly, it is not necessary to have a regular lattice for mean field theory to be valid, and having a relatively large average degree in the imprinted networks is sufficient.

Answering the second question, we found that the emergent network's measures depend only in a simple manner on the strength and direction of projective measurement, and that this dependence completely went away when we rescaled the curves by a single scaling factor. Thus, we showed that emergent mutual information networks for the transverse Ising model, unexpectedly, do \emph{not} share the robustness properties of classical complex networks: (1) the choice of network structure does not help with decoherence (attack); and (2) targeted decoherence (attack) is no more effective than random decoherence. We found the same conclusion also for imprinted networks with small $Z$, even though they lie outside the region of validity of mean-field theory. However, we cannot exclude the possibility that this lack of robustness is due to the small network sizes in our calculations. Excitingly, current quantum computing and quantum simulator experiments can realize the large network sizes required to conclusively demonstrate whether or not imprinted quantum complex networks are robust to decoherence.

Our work introduces a new toolkit to represent quantum correlations in strongly correlated systems with the emergent mutual information network, and to visualize the effects of decoherence by analyzing the emergent network. Our work also highlights important and unforeseen differences between classical and quantum networks. A better understanding of strongly correlated systems and decoherence using novel tools such as network analysis will help us design robust complex quantum systems, and ultimately, a robust quantum Internet.

\section*{Acknowledgment}
This work was performed in part with support by the NSF under grants CCF-1839232, OAC-1740130, PHY-1806372, and PHY-1748958; and in conjunction with the QSUM program, which is supported by the Engineering and Physical Sciences Research Council grant EP/P01058X/1. This work has been supported by the European Research Council under the Consolidator Grant COQCOoN (Grant No. 820079).
 B.S. acknowledges support from . M.W. and V.P. acknowledge support from .
 
The authors acknowledge Colorado School of Mines supercomputing resources~\cite{minesHPC} made available for conducting the research reported in this paper. We thank Sabrina Maniscalco for useful discussions.

\section*{References}
\bibliography{refs}

\appendix
\section{Node deletions on the imprinted networks}\label{sec: appendix A}

\begin{figure}[t]\centering
\includegraphics[width=1.0\columnwidth]{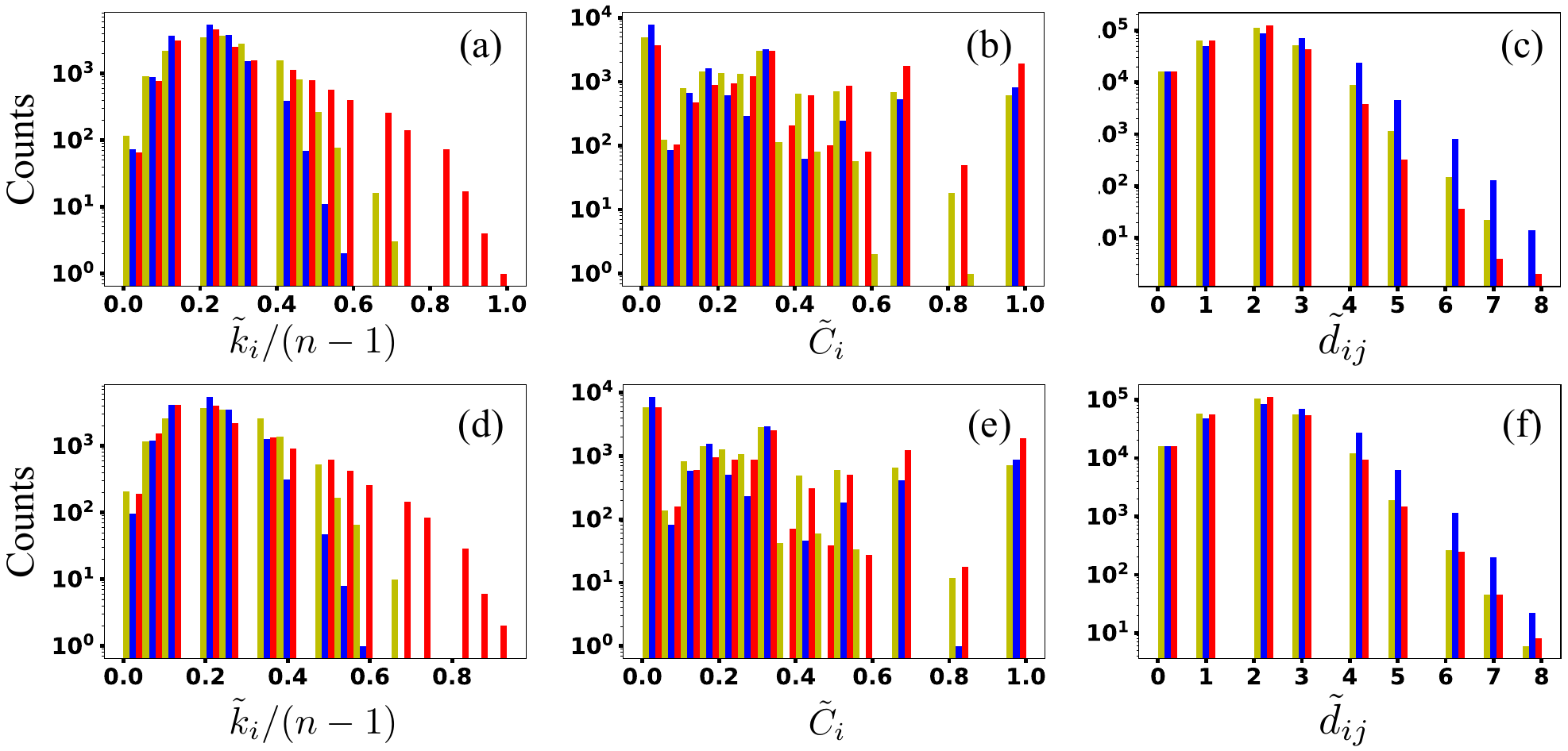}
\caption{\textit{Network measures after deletion of nodes.} (a-c) randomly remove $20\%$ of the nodes from each network with size $n=20$. (d-f) remove $20\%$ of nodes in a targeted manner. Targeted deletions in BA networks lead to fewer hubs, visible as slightly fewer nodes with a large degree. The path lengths are longer compared to the distribution before attack [see Fig.~\ref{fig1}(f)]. The parameters used to construct these networks are the same as those used in the main text.
}
\label{fig: supp1}
\end{figure}

\begin{figure}[t]\centering
\includegraphics[width=1.0\columnwidth]{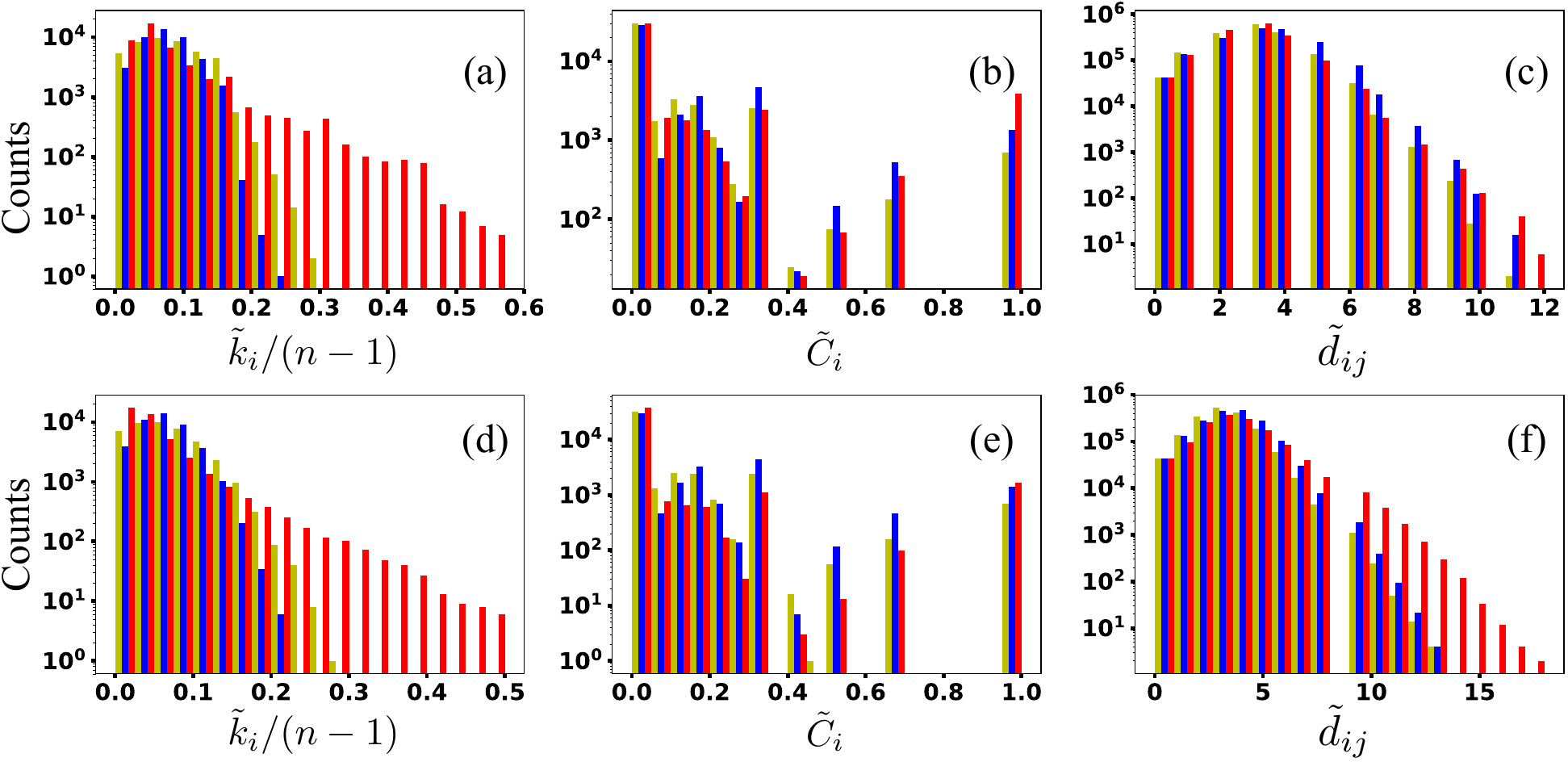}
\caption{\textit{Network measures after deletion of nodes.} (a-c) randomly remove $20\%$ of the nodes from each network with size $n=54$. (d-f) remove $20\%$ of nodes in a targeted manner. Removal of hubs by targeted deletions in BA is much more prominent for $n=54$ than for $n=20$. The ER networks have $p=0.04$, the WS networks have $K=4$ and $p=0.5$, and the BA networks have $m=2$.
}
\label{fig: supp2}
\end{figure}

Here, we consider an ensemble of $1000$ networks with $n=20$ and $n=54$. The latter number is motivated by Google's Sycamore chip~\cite{arute2019quantum}. We delete nodes either randomly or in a targeted manner, and plot the measures after node deletion.

Figures~\ref{fig: supp1}(a-c) shows the measures after $20\%$ of the nodes are randomly deleted from each network of size $n=20$, and Figs.~\ref{fig: supp1}(d-f) show the measures after deletion when the deleted nodes are targeted. The most prominent difference between the two cases is that when attacks are targeted, nodes with a large degree in the BA networks are fewer. This is because targeted deletions preferentially removes hubs, which tend to have a large degree. The distribution of path lengths after attacks has an exponential tail, and some path lengths are longer compared to Fig.~\ref{fig1}. This is because deleting nodes removes shortest paths for some node pairs.

Figures~\ref{fig: supp2}(a-c) shows the measures after $20\%$ of the nodes are randomly deleted from each network of size $n=54$, and Figs.~\ref{fig: supp2}(d-f) show the measures after deletion when the attacks are targeted. The effect of targeted deletions from BA networks is more visible here, especially in the number of nodes with a large degree and in the longer path lengths.

\section{Projective measurements along $\hz$}\label{sec: appendix B}

\begin{figure}[t]
\centering
\includegraphics[width=0.6\columnwidth]{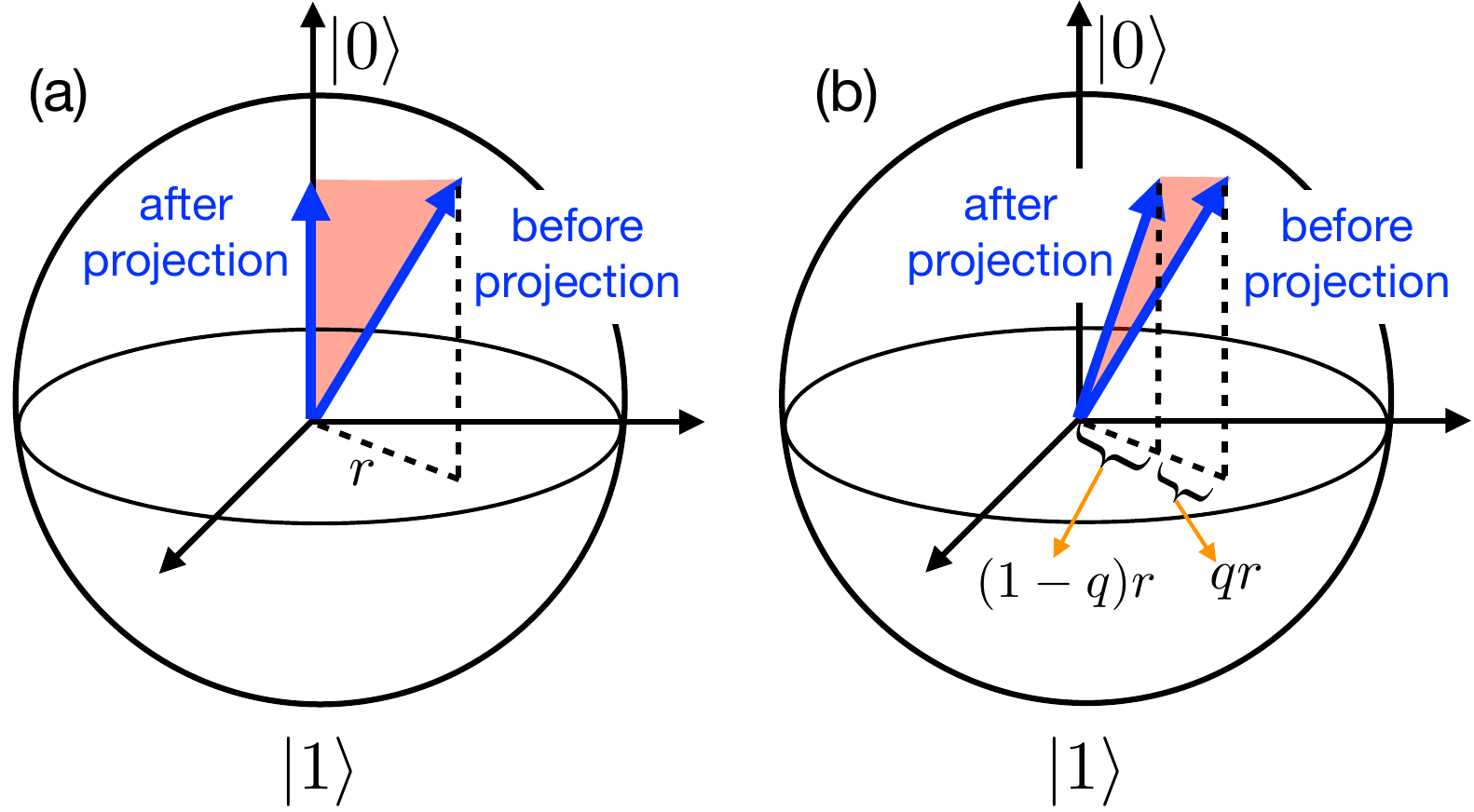}
\caption{\textit{Projective measurement attack on a spin}. (a) A complete projective measurement projects the spin's Bloch vector onto the axis of measurement (here, $\hz$), reducing its component on the plane orthogonal to measurement (here, the $x$-$y$ plane), from an initial value of $r$ to a final value of $0$. (b) A partial projective measurement reduces its component on this plane to $(1-q)r$.}
\label{fig: bloch sphere Z}
\end{figure}

\begin{figure}[t]
\centering
\includegraphics[width=0.95\columnwidth]{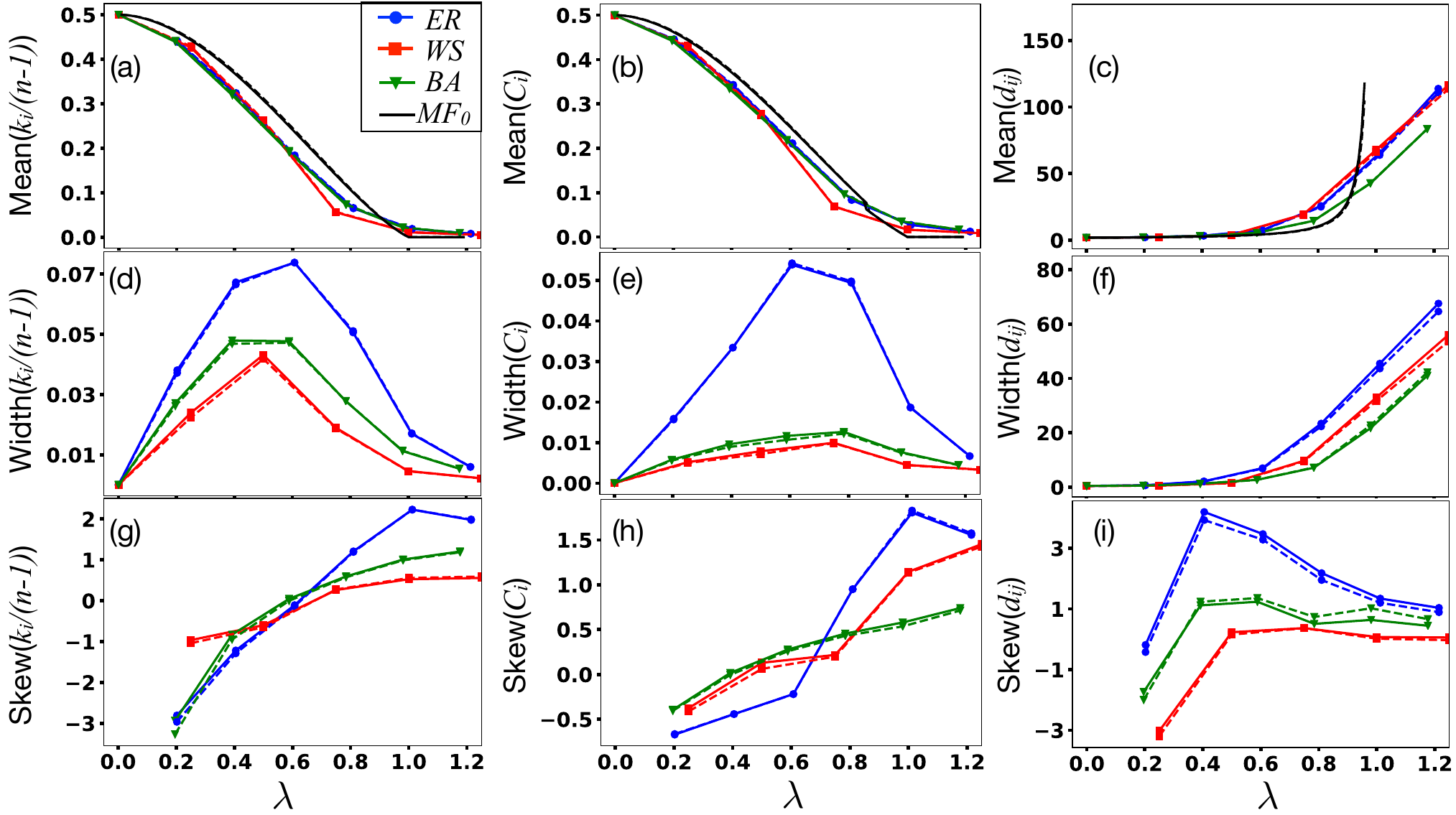}
\caption{\textit{Moments of the distributions of the measures $k_i$, $C_i$, and $d_{ij}$ in the emergent network, after $20\%$ of the nodes in the ground state of each imprinted network are randomly picked and projectively measured along $\hz$.} The different panels plot the same quantities as Fig.~\ref{fig6} with the same color coding. The moments of the measures are similar for both measurement strengths, and the mean values are captured well by MF$_0$.
}
\label{fig5}
\end{figure}

\begin{figure}[b]
\centering
\includegraphics[width=0.95\columnwidth]{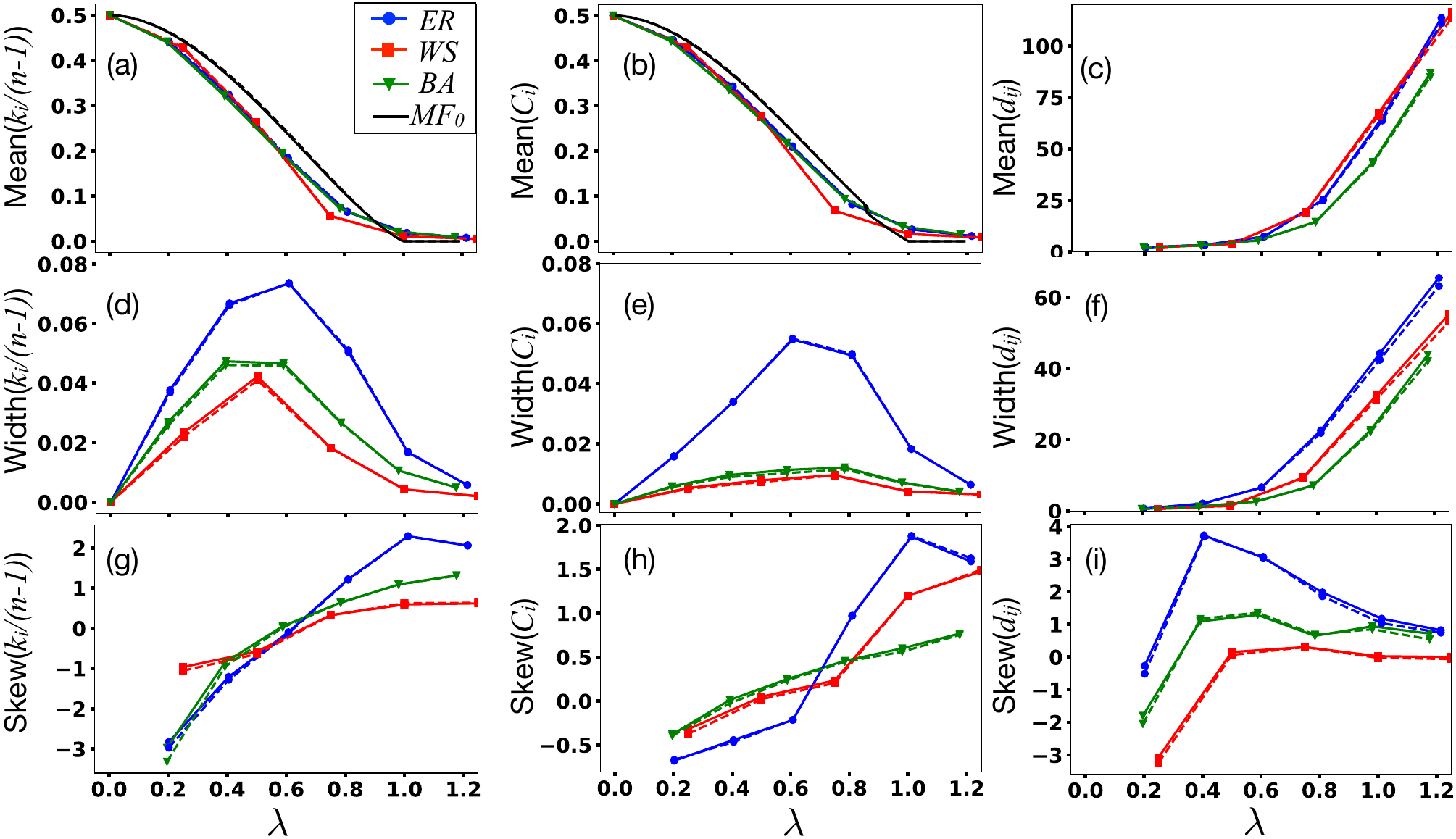}
\caption{ \textit{Moments of the distributions of the measures $k_i$, $C_i$, and $d_{ij}$ in the emergent network, after $20\%$ of the nodes in the ground state of each imprinted network are preferentially picked and projected onto $\hz$.} The different panels plot the same quantities as Fig.~\ref{fig6} with the same color coding. There is almost no difference between the results for targeted attacks (this figure) and random attacks [Fig.~\ref{fig5}].
}
\label{fig7}
\end{figure}

Complementary to Sec.~\ref{subsec: projective}, here we consider projective measurements along $\hz$, as shown in Fig.~\ref{fig: bloch sphere Z}. In Fig.~\ref{fig5}, we choose the attacked spins are chosen randomly, and in Fig.~\ref{fig7}, we choose them in a targeted manner. As in Figs.~\ref{fig6}-\ref{fig8}, panels (a-c) have eight curves each, and panels (d-i) six curves each, with the same color scheme.

Remarkably, we find little difference between the moments in Figs.~\ref{fig5} after the attacks, and the moments in Fig.~\ref{fig4} before the attacks. There is also almost no difference between the curves for $q=1$ and $q=0.5$. Moreover, we also find no difference in the results between the different strategies of choosing the projected nodes in Figs.~\ref{fig5}-\ref{fig7}. Thus, we find the surprising result that measuring along $\hz$ has no effect on the low order moments of $k_i/(n-1)$, $C-i$, and $d_{ij}$, regardless of the strength and strategy of attack. Mean field theory (MF$_0$) accurately captures the mean values of these measures.

\section{Mean field theory for projective measurements along $\hz$}
When a node $i$ is attacked with a projective measurement along $\hz$, its magnetization $\braket{\hs^x_i}$ and all correlations $\braket{\hs^x_i\hs^x_j}$ get shrunk by $(1-q)$. The node's single-particle density matrix in MF theory after projection is
\begin{equation}
\rho_i = \left( \begin{array}{cc} \frac{1}{2} & \frac{(1-q)\sqrt{1-m^2}}{2} \\ \frac{(1-q)\sqrt{1-m^2}}{2} & \frac{1}{2} \end{array}\right).
\end{equation}
Similarly, the two-particle density matrix after an attack on one node becomes
\begin{equation}
\hspace*{-2.5cm}
\rho_{ij} = \frac{1}{4} \left( \begin{array}{cccc} 
1+m^2 & \sqrt{1-m^2} & (1-q)\sqrt{1-m^2} & (1-q)(1-m^2) \\
\sqrt{1-m^2} & 1-m^2 & (1-q)(1-m^2) & (1-q)\sqrt{1-m^2} \\
(1-q)\sqrt{1-m^2} & (1-q)(1-m^2) & 1-m^2 & \sqrt{1-m^2} \\
(1-q)(1-m^2) & (1-q)\sqrt{1-m^2} & \sqrt{1-m^2} & 1+m^2
\end{array}\right).
\end{equation}
The two-particle density matrix after an attack on both nodes becomes
\begin{equation}
\hspace*{-2.5cm}
\rho_{ij} = \frac{1}{4} \left( \begin{array}{cccc} 
1+m^2 & (1-q)\sqrt{1-m^2} & (1-q)\sqrt{1-m^2} & (1-q)^2(1-m^2) \\
(1-q)\sqrt{1-m^2} & 1-m^2 & (1-q)^2(1-m^2) & (1-q)\sqrt{1-m^2} \\
(1-q)\sqrt{1-m^2} & (1-q)^2(1-m^2) & 1-m^2 & (1-q)\sqrt{1-m^2} \\
(1-q)^2(1-m^2) & (1-q)\sqrt{1-m^2} & (1-q)\sqrt{1-m^2} & 1+m^2
\end{array}\right).
\end{equation}

\end{document}